\documentclass[10pt,natbib,twocolumn,showpacs,preprintnumbers,amsmath,amssymb]{revtex4}

\usepackage{graphicx}  
\usepackage{dcolumn}   
\usepackage{bm}        

\newcommand\Eqn[1]     {Eq.\,(\ref{#1})}
\newcommand\Eqns[2]    {Eqs\,(\ref{#1}) and~(\ref{#2})}

\newcommand\Fig[1]     {Fig.\,{\ref{#1}}}

\newcommand\nn         {\nonumber}

\newcommand{\be}{\begin{equation}}
\newcommand{\ee}{\end{equation}}
\newcommand{\ba}{\begin{eqnarray}}
\newcommand{\ea}{\end{eqnarray}}

\def\Lin{{\rm Lin}}

\def\pp1{{\prime}}
\def\pp2{{\prime\prime}}

\def\2D{{\rm 2D}}

\def\Vu{{V_{\mu}}}

\def\cyc{{\rm cyc}}

\def\h{{\rm h}}

\def\bx{{\bf x}}
\def\br{{\bf r}}

\def\bk{{\bf k}}
\def\bq{{\bf q}}
\def\by{{\bf y}}

\def\cent{{\rm cent}}
\def\1Loop{{\rm 1Loop}}

\def\rhob{\bar{\rho}}

\def\Msol{h^{-1}M_{\odot}}
\def\kpc{\, h^{-1}{\rm kpc}}
\def\Mpc{\, h^{-1}{\rm Mpc}}
\def\Mpccube{\, h^{-3} \, {\rm Mpc}^3}

\def\Gpccube{\, h^{-3} \, {\rm Gpc}^3}
\def\kMpc{\, h \, {\rm Mpc}^{-1}}

\def\hh{{\rm hh}}

\def\dr{{\rm d}^3{\bf r}}

\def\dk{{\rm d}^3{\bf k}}
\def\dq{{\rm d}^3{\bf q}}
\def\dy{{\rm d}^3{\bf y}}

\def\nbar{\bar{n}}

\font\BF=cmmib10

\def\by{{\hbox{\BF y}}}

\def\fun#1#2{\lower3.6pt\vbox{\baselineskip0pt\lineskip.9pt
        \ialign{$\mathsurround=0pt#1\hfill##\hfil$\crcr#2\crcr\sim\crcr}}}

\def\FNL{{f_{\rm NL}}}
\def\hc{{\rm h}}
\def\1H{1{\rm H}}
\def\2H{2{\rm H}}
\def\3H{3{\rm H}}

\def\vir{{\rm vir}}
\def\phiG{{\rm \phi}}
\def\phiNG{{\rm \Phi}}
\def\PphiG{P_{\phiG}}
\def\PphiNG{P_{\phiNG}}
\def\funk{{\mathcal F}_{\left\{M,R\right\}}}

\begin{document}


\title{Nonlinear clustering in models with primordial non-Gaussianity:\\
  the halo model approach}


\author{Robert E. Smith$^{1,2}$, Vincent Desjacques$^{1}$ \&  Laura Marian$^{2}$}

\affiliation{
\vspace{0.2cm}
(1) Institute for Theoretical Physics, University of Zurich, Zurich CH 8037, Switzerland\\
(2) Argelander-Institute for Astronomy, Auf dem H\"ugel 71, D-53121 Bonn, Germany} 

\email{res@physik.unizh.ch}


\begin{abstract}
  We develop the halo model of large-scale structure as an accurate
  tool for probing primordial non-Gaussianity (PNG). In this study we
  focus on understanding the matter clustering at several redshifts in
  the context of PNG that is a quadratic correction to the local
  Gaussian potential, characterized by the parameter $\FNL$. In our
  formulation of the halo model we pay special attention to the effect
  of halo exclusion, and show that this can potentially solve the long
  standing problem of excess power on large scales in this model.  The
  halo model depends on the mass function, clustering of halo centers
  and the density profiles.  We test these ingredients using a large
  ensemble of high-resolution Gaussian and non-Gaussian numerical
  simulations, covering $\FNL=\{0,+100,-100\}$. In particular, we
  provide a first exploration of how halo density profiles change in
  the presence of PNG. We find that for $\FNL$ positive/negative high
  mass haloes have an increased/decreased core density, so being
  more/less concentrated than in the Gaussian case. We also examine
  the halo bias and show that, if the halo model is correct, then
  there is a small asymmetry in the scale-dependence of the bias on
  very large scales, which arises because the Gaussian bias must be
  renormalized. We show that the matter power spectrum is modified by
  $\sim2.5\%$ and $\sim3.5\%$ on scales $k\sim1.0\kMpc$ at $z=0.0$ and
  $z=1.0$, respectively. Our halo model calculation reproduces the
  absolute amplitude to within $\lesssim10\%$ and the ratio of
  non-Gaussian to Gaussian spectra to within $\lesssim1\%$. We also
  measure the matter correlation function and find similarly good
  levels of agreement between the halo model and the data. We
  anticipate that this modeling will be useful for constraining $\FNL$
  from measurements of the shear correlation function in future weak
  lensing surveys such as Euclid.
\end{abstract}




\maketitle


\section{Introduction}\label{sec:intro}

Over the last few decades, through experiments such as 2dFGRS, SDSS
and WMAP \citep{Collessetal2003,Abazajianetal2009,Jarosiketal2010},
great strides have been made in quantifying the parameters of the
perturbed and unperturbed Friedmann-Lema\^itre-Robertson-Walker (FLRW)
model. Besides the detailed information about the present Universe,
these experiments have also opened up important new windows into the
physics of the early Universe.

The inflationary paradigm is so far the leading physical explanation
for the origins of structure. The single-field slow-roll theory makes
four fundamental predictions: a flat universe (quantified by the
curvature density parameter $\Omega_K$); a primordial density power
spectrum with power-law index ($n_s$) just less than unity; a nearly
Gaussian distribution of primordial density fluctuations (deviation
from Gaussianity being quantified by the $\FNL$ parameter); a spectrum
of gravitational waves (characterized by the amplitude of the
quadrupole tensor to scalar ratio $r$). The measurement of the CMB
temperature anisotropies combined with a number of cosmological
probes, such as the Baryonic Acoustic Oscillation (BAO) scale in the
SDSS LRGs, and the present day value of the Hubble constant, provide
strong supporting evidence in favor of the first three of these
predictions. The current WMAP7+BAO+$H_0$ combined constraints on these
parameters are \citep{Komatsuetal2010}: $-0.0133<\Omega_{K}<0.0084$
(95\% CL); $n_s=0.963\pm0.012$ (68\% CL); $\FNL=32\pm21$ (68\% CL).
For the spectrum of gravitational waves current CMB experiments
(WMAP7+ACBAR+QUaD) place an upper bound of $r<0.33$ (95\% CL)
\cite{Komatsuetal2010,Reichardtetal2009,Brownetal2009}. The latter can
only be further falsified with dedicated CMB polarization experiments
such as PLANCK\cite{Planck2006}.

Intriguingly, the constraints on the amount of primordial
non-Gaussianity (hereafter, PNG) from the CMB are tightening around a
non-zero value, there currently being $\sim1.5\sigma$ evidence against
pure Gaussian scalar perturbations \citep{Komatsuetal2010} (see also
\citep{YadavWandelt2008} who claimed a 99.5\% CL detection of non-zero
$\FNL$). If $\FNL$ is found to be substantially greater than zero at
high significance, then this would rule out {\em all} inflation models
based on a single scalar field \citep{CreminelliZaldarriaga2004}. On
the other hand, multi-field models may produce large levels of PNG and
also scale-dependent $\FNL$. Thus testing the Gaussianity of the
initial fluctuations is of prime concern.

Most tests for PNG have primarily been performed on the temperature
anisotropies in the CMB, however for several decades it has been
theoretically understood that PNG also affects a number of large-scale
structure observables
\citep{GrinsteinWise1986,FryScherrer1994,LucchinMatarrese1988,Matarreseetal2000}.
However, very few tests with real galaxy survey data have been
performed.  This was primarily due to the fact that in the present day
Universe the density fluctuations do not remain pristine, but have
been driven to a non-Gaussian state by gravity. Gravitational
evolution of the density perturbations correlates the amplitudes and
phases of different Fourier modes, thus one is faced with the problem
of decoupling primordial from gravitational
non-Gaussianity. Furthermore, in observing large-scale structures one
does not in general get information about all points in space, but
instead one is restricted to learning only about the galaxy
distribution. In general this is related to the underlying density
statistics through a bias function (or functional), which may be
complicated and stochastic. In the simple case of deterministic local
biasing, one may attempt to use higher order statistics such as the
galaxy bispectrum to disentangle the effects of gravity and galaxy
bias
\citep{Scoccimarroetal2004,SefusattiKomatsu2007,Nishimichietal2010,Sefusattietal2010}.
However, no constraints on $\FNL$ have yet emerged from such
schemes. This partly owes to the fact that past survey volumes have
been too small for such tests to be performed with any confidence. The
survey volumes of ongoing and future planned missions will surely
change this.

Recently, in a ground breaking paper, it was shown both theoretically
and in numerical simulations by \citep{Dalaletal2008}, that there is a
strong signature of $\FNL$ in the power spectrum of dark matter
haloes. The effect is to induce a scale-dependent bias correction
$\Delta b\propto k^{-2}$. The exciting prospect of this is that, since
the signature affects primarily only the largest scales,
$k<0.02\kMpc$, one can in principle decouple the effects induced by
nonlinear bias and gravity from those of PNG, and so constrain the
latter. There has been much activity in quantifying the effects of
this scale-dependent bias on the power spectrum in simulations
\citep{Desjacquesetal2009,Grossietal2009,Pillepichetal2010}; and also
there has been much theoretical activity to truly understand how the
scale-dependent bias arises
\citep{Slosaretal2008,MatarreseVerde2008,Taruyaetal2008,Sefusatti2009,GiannantonioPorciani2010},
and for a current review see \citep{DesjacquesSeljak2010}.  This has
culminated in several recent attempts to constrain $\FNL$ from
large-scale structure data
\citep{Slosaretal2008,Xiaetal2010,DeBernardisetal2010}. 

Another recent result has shown that the nonlinear dark matter power
spectrum alone is sensitive to the presence of $\FNL$
\citep{Grossietal2008,Desjacquesetal2009,Pillepichetal2010,Nishimichietal2010,Sefusattietal2010}.
This prompted \citep{FedeliMoscardini2010} to propose that the
statistics of the initial conditions could be tested through weak
gravitational lensing by large-scale structure, in particular through
measurement of the two-point shear correlation functions.  The
advantage of such a probe is that it is only sensitive to the total
mass distribution projected along the line of sight. On the down side,
the signal is weak, with deviations being of the order several
percent. However, future all sky weak lensing surveys such as EUCLID
\citep{EUCLID2010} and LSST \citep{LSST2009} will be able to probe
changes in the convergence power spectrum to percent level accuracy.
It is therefore of great interest to quantify in detail how sensitive
such a mission would be to constraining PNG of the local type. The
starting point for such a study is an accurate model of the matter
power spectrum as a function of redshift. 

In this study we focus on understanding how PNG shapes the two-point
matter clustering statistics in Fourier and configuration
space. Currently there is no accurate analytic model for describing
the effects of PNG on the matter power spectrum over the range of
scales that will be required for future weak lensing missions
($k\in[0.01,100.0]\kMpc$
\citep[][]{HutererTakada2005,Hutereretal2006}). This we shall attempt
to build.  Following \citep{AmaraRefregier2004},
\citep{FedeliMoscardini2010} proposed that the halo model might be
able to predict the power spectrum. They suggested that PNG would
modify predictions in two ways: through the abundance of massive
clusters; and through the scale-dependent bias. No attempt to compare
their model with numerical simulations was presented and so it remains
an open question as to how reliable this proposition is. Moreover,
they assumed that halo profiles are not affected by PNG. In this work
we show, through detailed numerical simulations, that this assumption
is wrong, and that there are important effects which need to be taken
into account if the halo model is to be used to make accurate
predictions.

Furthermore, we make important new developments in the importance of
halo exclusion in order to make robust predictions, and we show for
the first time how this may also resolve a long-standing problem with
the halo model, the excess power on large scales. 

Lastly, when exploring how the scale-dependent bias from PNG enters
the halo model framework, we have shown that if the model is to be
self-consistent, then there must be a modification to the original
formula of \citep{Dalaletal2008}.  This modification creates an
asymmetry in the bias for models with the same value of $\FNL$ but
with an opposite sign. This arises due to the fact that it is not the
Gaussian bias which enters the formula of Dala, but in fact the total
non-Gaussian scale-independet bias. We believe that inclusion of this
effect will slightly modify current constraints on $\FNL$.

The paper is broken down as follows: In \S\ref{sec:ng} we overview the
local model for PNG and its impact on density statistics. In
\S\ref{sec:simulations} we describe the suite of $N$-body simulations
used in this study. In \S\ref{sec:modelling} we explore modeling of
the matter power spectrum with PNG, in particular first explore the
perturbation theory approach, before moving on to describing the halo
model.  In \S\ref{sec:ingred} we perform a phenomenological study of
the necessary ingredients of the halo model and their dependence on
PNG. In \S\ref{sec:results} we present our predictions for the halo
model power spectrum as a function of redshift and also the two point
correlation function. Finally, in \S\ref{sec:conclusions} we summarize
our findings and conclude.


\section{Primordial non-Gaussianity}\label{sec:ng}

\subsection{Potential statistics}

We shall be working with the local model of PNG, that is we consider
only local quadratic corrections to the Gauge invariant Bardeen's
potential perturbation, which on scales smaller than the horizon size
reduces to minus the Newtonian potential \citep{Matarreseetal2000}:
\be 
\phiNG(\bx)=\phiG(\bx)+\FNL \left[\phiG(\bx)^2-\left<\phiG^2(\bx)\right>\right]\ ,
\label{eq:fnl}
\ee
where $\phiG(x)$ is the Gaussian potential perturbation after matter
radiation equality, scaled in terms of units of $c^2$ to yield a
dimensionless quantity. Following standard convention, it is defined
to be related to the Newtonian potential as
$\phiNG(\bx)\equiv-\Phi^{\rm Newton}(\bx)$. The term
$\left<\phiG^2(x)\right>$ is subtracted to ensure that $\phiNG$ is a
mean zero field. In linear theory the typical fluctuations are of the
order $\phiNG\sim10^{-5}$, and so the non-Gaussian corrections are of
the order $\sim 0.1\% (\FNL/100)(\phi/10^{-5})$.

This transformation of the Gaussian potential leads to a small
correction to the power spectrum, but its main effect is to generate a
primordial potential bispectrum. To see this consider the Fourier
transform of \Eqn{eq:fnl}:
\be 
\phiNG(\bk)=\phiG(\bk)+\FNL \int \frac{\dq}{(2\pi)^3}
\phiG(\bq)\phiG(\bk-\bq) .
\ee
The power spectrum, defined as
$\left<\phiNG(\bk_1)\phiNG(\bk_2)\right> \equiv
P_{\phiNG}(\left|\bk_1\right|)(2\pi)^3\delta^D(\bk_1+\bk_2)$ for this field is given by:
\be 
\PphiNG(\bk_1) = \PphiG(\bk_1)
+2\FNL^2\int\frac{\dq_1}{(2\pi)^3}\PphiG(\bq_1)\PphiG(\left|\bk_1-\bq_1\right|)\ .
\label{eq:potpow}
\ee
The three point function in Fourier space is given by
\ba 
\left<\phiNG(\bk_1)\phiNG(\bk_2)\phiNG(\bk_3)\right> 
& = & \nn \\
&  & \hspace{-4cm}
\FNL \int \frac{\dq_3}{(2\pi)^3} \left<\phi(\bk_1)\phi(\bk_2)
\phi(\bq_3)\phi(\bk_3-\bq_3)\right>+2\cyc \nn \\
&  & \hspace{-4cm} + \FNL^3 
\int \frac{\dq_1}{(2\pi)^3}\frac{\dq_2}{(2\pi)^3}\frac{\dq_3}{(2\pi)^3} 
\left<\frac{}{}\phi(\bq_1)\phi(\bk_1-\bq_1)
\right. \nn \\
&  & \hspace{-3cm} \times \left. \frac{}{}
\phi(\bq_2)\phi(\bk_2-\bq_2)\phi(\bq_3)\phi(\bk_3-\bq_3)
\right>\ \label{eq:primebi}.
\ea
Recalling that the expectation value of odd powers of the Gaussian
variables vanish, and from Wick's theorem we have that the even
powers can be written as
$\left<\phi(\bk_1)\dots\phi(\bk_n)\right>=\sum_{\rm all pairs}
\prod_{i=1}^{\rm pairs} \left<\phi(\bk_i)\phi(-\bk_i)\right>$.
Also, defining the bispectrum as
$\left<\phiNG(\bk_1)\phiNG(\bk_2)\phiNG(\bk_3)\right>\equiv
B_{\phiNG}(\bk_1,\bk_2,\bk_3)(2\pi)^3\delta^D(\bk_1+\bk_2+\bk_3)$,
then we find the primordial potential bispectrum to be:
\ba 
B_{\phiNG}(\bk_1,\bk_2,\bk_3)
& = &
2\FNL 
\left[
\PphiG(\bk_1) \PphiG(\bk_2) + 2 \, {\rm cyc}\ \right]\nn\\
& + &  4 \FNL^3 \int \frac{\dq_1}{(2\pi)^3}\PphiG(\bq_1)\PphiG(\left|\bq_1-\bk_1\right|)\nn\\
& \times & \left[\frac{}{}\PphiG(\left|\bq_1+\bk_2\right|)+\PphiG(\left|\bq_1+\bk_3\right|)\right]
\ .\label{eq:potbi}\ea
Restricting our attention to the case where $\FNL<100$ then we may
safely neglect the second terms on the right-hand-side of
\Eqns{eq:potpow}{eq:potbi} \footnote{At first glance the nonlinear (in
  $\FNL$) corrections to the power and bispectrum in
  \Eqns{eq:potpow}{eq:potbi}, appear to be infra-red divergent
  quantities. This can be seen most simply by taking the limit
  $k\rightarrow0$, and for the power spectrum this leads to an
  integral of the form $\int_{0}^{\infty} dq q^2
  P_{\phi}^2(q)$. Taking $P_{\phi}(q) \propto k^{n-4}$, then the above
  integral becomes $\int_{0}^{\infty} dq q^{2(n-4)+2}\propto
  \left.k^{\alpha}\right|_0^{\infty}$, where $\alpha=2(n-4)+3$. Taking
  $n=1$, then we get $\alpha=-3$, and we now see that if we substitute
  the lower limit then the expression diverges. However, as was
  pointed out by \citep{GiannantonioPorciani2010}, if one believes
  that inflation was responsible for the non-Gaussianity, then one
  should invoke the cut-off scales which limit the physical processes:
  on small scales the fluctuations are cut-off by the reheating scale
  during inflation, and on the large scales, by the horizon size
  today. In this case and for realistic values of $\FNL$, then the
  higher order corrections should be subdominant at the present time,
  and so can be neglected
  \cite{Matarreseetal2000,GiannantonioPorciani2010}.}


\subsection{Density statistics}

The primordial matter potential and density fluctuations, extrapolated
to the present day, can be related through Poisson's equation:
\ba
\nabla^2\phiNG(\bx,a) & = & 
- \frac{4\pi G}{c^2} \left[\rho(\bx,a)-\rhob(a)\right]a^2 \nn \\ 
& = &
-\frac{3}{2}\Omega_0\left(\frac{H_0}{c}\right)^2  \frac{D(a)}{a}\delta_0(\bx,a_0)\ ,
\ea
where $a$ is the expansion factor, $\rhob(a)=\Omega(a)\rho_{\rm
  crit}(a)\propto a^{-3}$ is the mean density of the Universe,
$\Omega(a)$ is density parameter, $\rho_{\rm crit}(a)$ the critical
density, $H(a)$ is the Hubble parameter, $D(a)$ is the linear growth
factor normalized to be unity at the present time, and quantities
labeled with a subscript 0 are to be evaluated at the present time
$a_0$. This equation may be solved in Fourier space to give an
explicit relation for the potential:
\be
\phiNG(\bk,a) = \frac{3}{2}\frac{\Omega_0}{k^2}
\left(\frac{H_0}{c}\right)^2  \left(\frac{D(a)}{D_0}\right)
\left(\frac{a_0}{a}\right)\delta_0(\bk,a_0)\ ,
\ee
Evolving the potential back to the initial epoch $a_i$, and dividing
the transfer function, then we have the following relation between the
present day density and primordial potential perturbations,
\be
\delta_0(\bk,a_0)  =  \alpha(k,a_i,a_0)\phiNG(\bk,a_i) \ ,
\label{eq:denpot}
\ee
where we have defined
\be 
\alpha(k,a_i,a_0)  \equiv  \frac{2 c^2 g(a_i,a_0)k^2 T(k,a_0)}{ 3\Omega_0 H_0^2}\ .
\label{eq:alpha}
\ee
In the above equation
$g(a_1,a_2)\equiv\left[D(a_2)/D(a_1)\right]\left[a_1/a_2\right]$
is the growth suppression factor ($a_1<a_2$), and for LCDM,
$g(a_i,a_0)\approx0.75$. 

In possession of the mapping from the present day density to
primordial potential perturbations through \Eqn{eq:denpot}, we may now
examine statistics of the density field. The most important statistic
that we will need to know is the present day skewness filtered on the
mass scale $M$. Following Appendix \ref{app:skew}, this can be
written:
\ba
\left<\delta^3_M(x,a_0)\right> 
& = & 6\FNL \int
\frac{dk_1}{2\pi^2}k_1^2W(k_1,M)\alpha(k_1)\PphiG(k_1) \nn \\ 
& & \times \ \int \frac{dk_2}{2\pi^2}k_2^2W(k_2,M)\alpha(k_2)\PphiG(k_2) \nn \\
& & \times \ \frac{1}{2}\int_{-1}^{1}d\mu W(k_3,M)\alpha(k_3) \ ,
\label{eq:skew}
\ea
where \mbox{$k_3^2\equiv k_1^2+k_2^2+2k_1k_2\mu$}, $W(k,M)$ is a
filter function that selects the mass scale $M$, and where for brevity
we shall make the definitions $\PphiG(k_2)\equiv \PphiG(k_2,a_i)$ and
$\alpha(k)\equiv \alpha(k,a_i,a_0)$. In what follows we shall also
make use of the reduced skewness, defined to be
\mbox{$S_3(M)\equiv\left<\delta_M^3(x)\right>/\left<\delta_M^2(x)\right>^2$}.
Finally, for completeness, we may write the density power spectrum at
some arbitrary epoch $a$ in terms of the primordial potential power
spectrum simply as:
\ba P_{\delta}(k,a) & = & D^2(a,a_0) \alpha^2(k,a_i,a_0) P_{\phi}(k,a_i) \nn \\
                   & = & D^2(a) \alpha^2(k) P_{\phi}(k) \ . \ea
%


\begin{figure}
\centering{ \includegraphics[width=7cm,clip=]{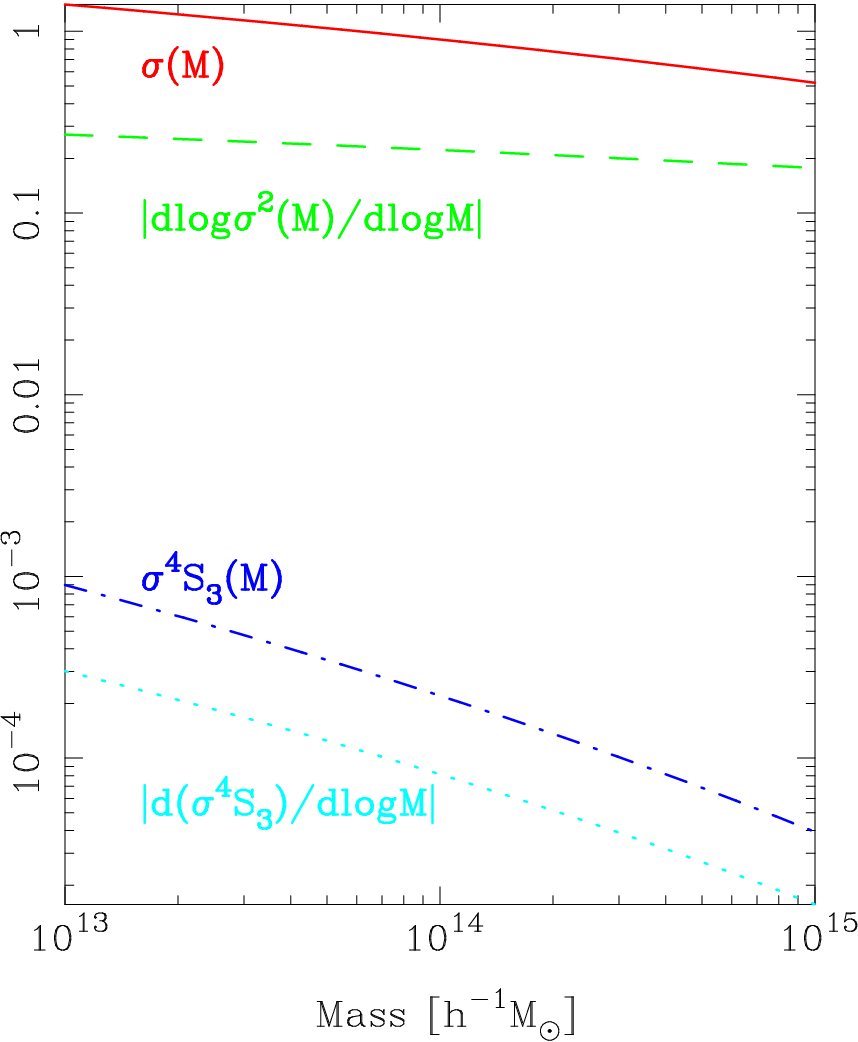}}
\caption{\small{Statistical description of the present day
    non-Gaussian density field, with $\FNL=1$, as a function of
    Lagrangian mass scale $M$. The solid line shows the standard
    deviation of the density field; the dot-dash line shows the
    skewness of the density field; the dash and dotted lines show the
    logarithmic derivatives of these quantities.}\label{fig:skewness}}
\end{figure}


In Figure \ref{fig:skewness} we show the present day skewness of the
density field $S_3(M)\sigma^4(M)$, as given by \Eqn{eq:skew} and with
$\FNL=1$. For the calculation we employ a real space spherical-top-hat
filter with radius given by the mass scale $R=(3M/4\pi\rhob)^{1/3}$,
which in Fourier space has the form: $W(y)= (3/y)\left[\sin y-y\cos
  y\right]$, where $y\equiv kR$. The figure shows that the skewness is
very small, $\lesssim10^{-3}$ for $R\sim5\Mpc$; and if $\FNL=100$, it
is $\lesssim0.1$.


\section{The $N$-body simulations}\label{sec:simulations}

In order to explore the impact of PNG on the clustering statistics of
the density field, we have generated a large ensemble of
high-resolution $N$-body simulations of the $\Lambda$CDM cosmology
seeded with Gaussian and non-Gaussian initial conditions.  This set is
an augmented version of the ensemble used by
\citet{Desjacquesetal2009} to study the mass function of CDM haloes
and their bias, and in \citet{Sefusattietal2010} to explore the matter
bispectrum.

The non-Gaussianity in the simulations is of the local form,
c.f.~\Eqn{eq:fnl}, and we use twelve sets of three simulations, each
of which has $\FNL=0,\pm 100$. Each simulation was run with $N=1024^3$
particles in boxes of side $L=1600\Mpc$ and $V_{\rm
  Sim}\sim4.096\Gpccube$, and this gives us a total simulated volume
of $V_{\rm Tot}\sim 49.152\Gpccube$. The interparticle forces were
softened on scales of 0.04 times the mean inter-particle distance,
which corresponds to $l_{\rm soft}\approx 40\kpc$.  We used the WMAP5
cosmological parameters: $\{h=0.7$, $\Omega_{\rm m}=0.279$,
$\Omega_{\rm b}=0.0462$, $n_s=0.96\}$, and a normalization of the
curvature perturbations $\Delta^2_{\cal R}(k)=2.21\times
10^{-9}(k/k_0)^{n_s-1}$, with $k=0.02$Mpc$^{-1}$, where the curvature
perturbation is related to the scalar potential ${\cal R}=5\Phi/3$,
and $\Phi$ was defined in \Eqn{eq:fnl}.  In terms of the variance of
matter fluctuations linearly extrapolated to the present day, this
gives $\sigma_8\approx 0.81$, where the variance is computed with a
spherical-top-hat filter of comoving radius $R=8\Mpc$.

The matter transfer function was generated using {\tt CAMB}
\cite{Lewisetal1999}.  All of the simulations were run using the
$N$-body code {\tt Gadget-2} \citep{Springel2005}. The same Gaussian
random seed field $\phi$ is employed for each $\FNL=\{0,+100,-100\}$
simulation set, and varied between sets. This allows the sampling
variance between different models of $\FNL$ to be minimized when we
construct statistics from the ratios of observables. The initial
particle distribution was generated at redshift $z_i=99$ using the
Zel'dovich approximation \citep{Zeldovich1970}.

Regarding the generation of the initial conditions for the non-Gaussian
simulations, we adopt the standard (CMB) convention in which
$\Phi(\bx)$ is primordial, and not extrapolated to present
epoch. Furthermore, we point out that the local transformation to the
potential given by \Eqn{eq:fnl} is performed before multiplication by
the matter transfer function $T(k)$.

Dark matter halo catalogs were generated for all snapshots of each
simulation using the Friends-of-Friends (FoF) algorithm
\citep{Davisetal1985}. We set the linking-length parameter to the
standard $b=0.2$, where $b$ is the fraction of the inter-particle
spacing. For this we used the fast parallel {\tt B-FoF} code, kindly
provided by V.~Springel. The minimum number of particles for which an
object was considered to be a bound halo, was set to 20
particles. With particle mass $m_{\rm p}\approx3.0\times
10^{11}\Msol$, this gave us a minimum host halo mass of $6\sim10^{12}
M_{\odot}/h$ \footnote{Note that we assume that haloes, which are
  found with FoF halo finders in numerical simulations, have a mass
  $M_{\rm FoF}$ and that this corresponds to
$M_{\rm FoF} \equiv M_{200}=4/3 \pi r_{200}^3 200 \bar{\rho}$,
i.e. the mean overdensity inside a sphere of radius $r_{200}$ is 200
times the mean background density. The only issue that one must be
careful of when dealing with halo phenomenology is that the same mass
convention is adopted for all semi-analytic models: mass function,
profiles and bias etc. Where there are clashes of definition, we are
careful to obtain mappings between these schemes to ensure consistency
following the method of \citet{SmithWatts2005}.}.


\begin{figure*}
\centering{
  \includegraphics[width=8.5cm,clip=]{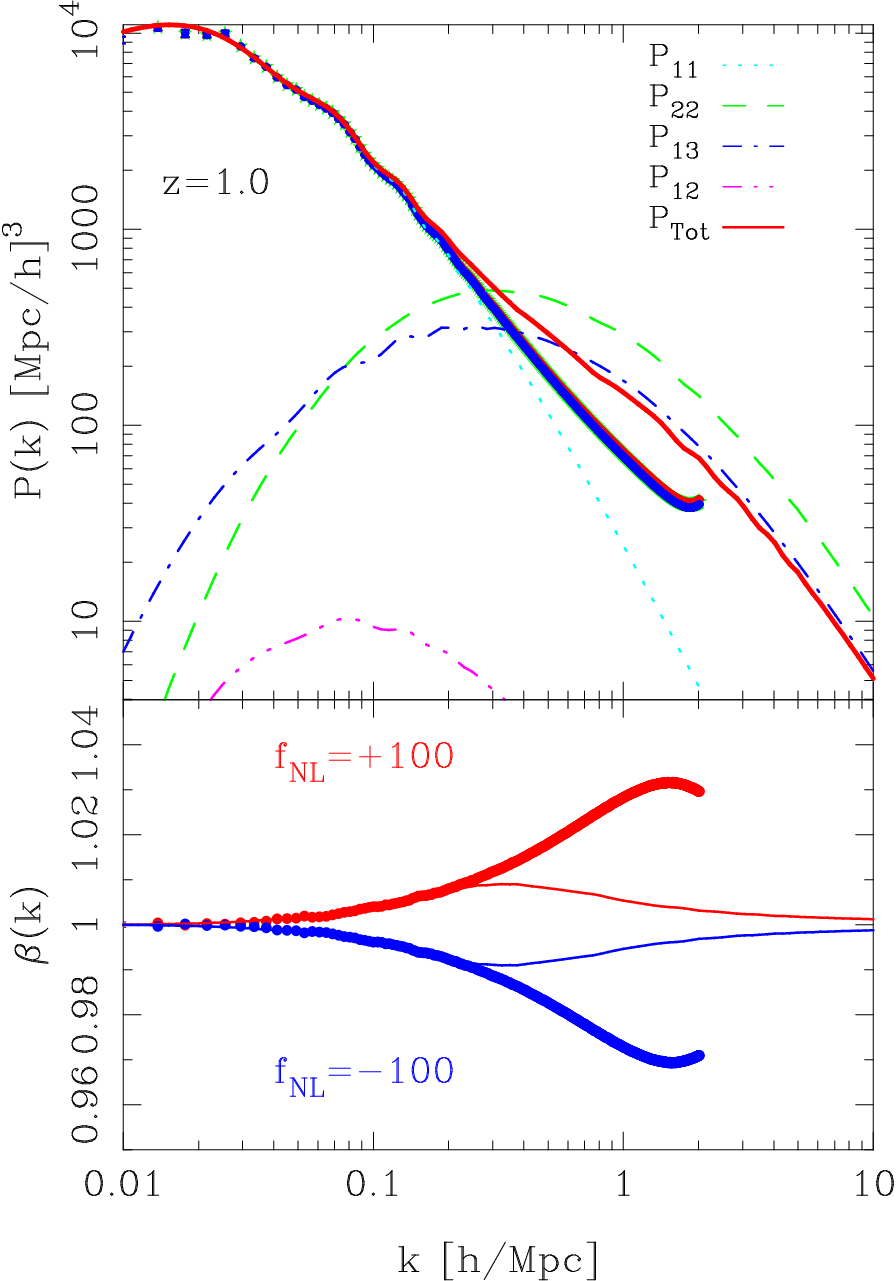}\hspace{0.2cm}
  \includegraphics[width=8.5cm,clip=]{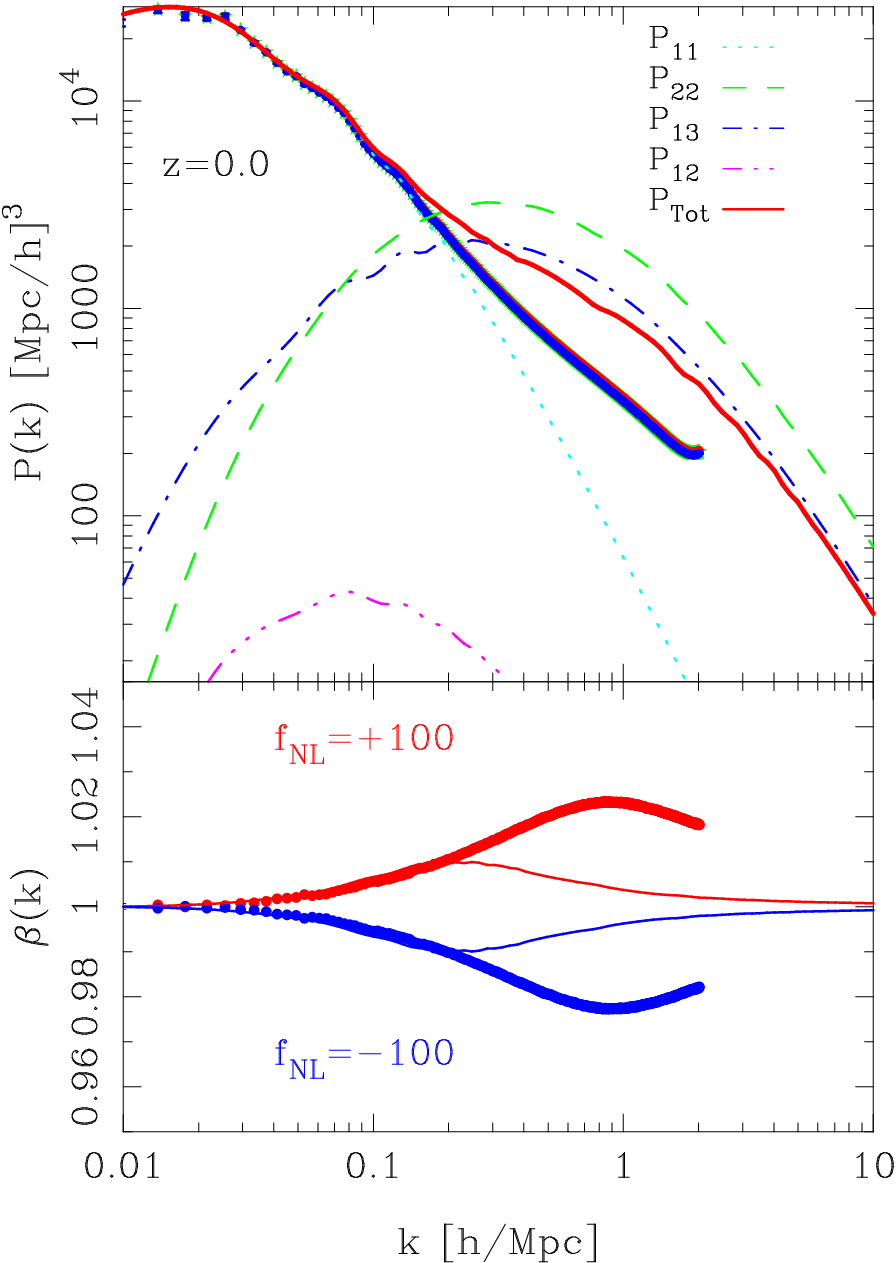}}
\caption{\small{Comparison of the nonlinear matter power spectra
    measured from the suite of $N$-body simulations with predicitons
    from nonlinear gravitational perturbation theory.  The left and
    right panels show results for $z=1.0$ and $z=0.0$,
    respectively. The top section of each panel presents the absolute
    power, the bottom section shows the ratio of the nonlinear power
    in the non-Gaussianity to the Gaussian models. Points with errors
    denote estimates from the simulations and lines denote theoretical
    predictions. The colors green, red and blue denote the models
    $\FNL=\{0,100,-100\}$, respectively. }\label{fig:PT}}
\end{figure*}


\section{Modeling the nonlinear power spectrum}\label{sec:modelling}

\subsection{Gravitational perturbation theory}

The nonlinear evolution of the density field can be followed using
gravitational perturbation theory. In this approach one writes down
the equations of motion for the CDM fluid in an expanding spacetime
and looks for a series expansion in the density and velocity
divergence of the CDM fluid \citep[for a review of the subject
  see][]{Bernardeauetal2002}. The important point to note is that the
perturbative solutions do not change when we consider the case of
structure formation in models with PNG. However, what does change is
the way in which the statistics of the density field behave. This owes
to the fact that there is now a complete hierarchy of connected
correlators of the field. Following the work of
\citep{Taruyaetal2008}, and keeping only terms that are linear in
$\FNL$, one finds that up to the next-to-leading-order corrections to
the matter power spectrum can be written:
\be P(k)=P_{11}(k)+P_{22}(k)+P_{13}(k)+P_{12}(k) \ ,\ee
where $P_{11}$, is the linear power spectrum, and $P_{13}$ and
$P_{22}$ are the so-called one-loop corrections, which appear in the
standard Gaussian theory \citep{Bernardeauetal2002}. The new term for
PNG is $P_{12}$, which has the form \citep{Taruyaetal2008}:
\ba P_{12}(k) & = & \frac{2 D^3(a) \FNL
  \alpha(k)P_{\phi}(k)k^3}
{7(2\pi)^2}\, \nn \\
& \times & \int^{y_{\rm max}}_{y_{\rm min}} \
\frac{dy}{y}y^3 \alpha(yk) P_{\phi}(yk) \nn \\
& \times & \,\int_{\rm max[-1,\epsilon_1]}^{\rm min[+1,\epsilon_2]}
d\mu \alpha(k\psi)\left(\frac{3y+7\mu-10y\mu^2}
  {1+y^2-2y\mu}\right)\ \nn \\
& \times & \left[1+\frac{P_\phi(k\psi)}{P_{\phi}(k)}+
  \frac{P_{\phi}(k\psi)}{P_{\phi}(yk)}\right] \;, \ea
where $\psi^2=k^2(1+y^2-2y\mu)$, $y_{\rm max}\equiv k_{\rm max}/k$ and
$y_{\rm min}\equiv k_{\rm min}/k$. We also defined 
\mbox{$\epsilon_1\equiv\left[k^2+q^2-k_{\rm max}^2\right]/2kq$}
and
\mbox{$\epsilon_2\equiv\left[k^2+q^2-k_{\rm min}^2\right]/2kq$}.
The cut-off scales are set to be $k_{\rm max}=10\kMpc$ and $k_{\rm
  min}=2\pi/L$, with $L$ being the simulation box length. Note that
the spectrum $P_{12}$ arises due to the existence of the non-zero
primordial potential bispectrum (c.f. Eqn.~\ref{eq:potbi}).

Let us now define the ratio of the power spectra in the non-Gaussian
and Gaussian models as:
\be \beta_{\rm PT}(k,\FNL)\equiv 1+
\frac{P_{12}(k,\FNL,a)}{P_{11}(k,a)+P_{13}(k,a)+P_{22}(k,a)}\ . \label{eq:beta}
\ee

In \citep{Desjacquesetal2009} it was shown that, for $k<0.2\kMpc$, the
PT description was able to capture to high precision the same ratio
measured from $N$-body simulations. In Figure~\ref{fig:PT} we extend
the analysis to much higher wavenumbers, relevant for cosmological
weak lensing studies.  On the left-hand-side of the figure, we show
the results for $z=1.0$ and on the right those for $z=0.0$ (we provide
further details of these measurements in \S\ref{sec:results}). The top
sections of the panels show the absolute power and on a log-log plot
the points for the three $\FNL$ models cannot be distinguished. The
bottom sections show $\beta_{\rm PT}(k,\FNL)$. We see that the effect
of $\FNL=\pm100$ is to induce $\pm\sim3.5\%$ modulations in the
nonlinear matter power spectrum at $z=1.0$ and on scales of order
$k=1.0\kMpc$, and that this reduces to $\sim2.5\%$ by $z=0.0$. These
deviations, although small would be larger than the measurement errors
in future weak lensing missions and so need to be accurately
characterized.

The figure also shows that, whilst PT captures well the behavior of
$\beta_{\rm PT}$ measured in the simulations on very large scales
$k<0.2\kMpc$, it fails to model the results on smaller scales. It is
also worth pointing out that whilst $\beta_{\rm PT}$ is well
characterized for $k<0.2\kMpc$, the absolute power does not match well
the measured absolute power at that scale, it being a factor of
$\sim2$ higher than the measurements at $z=0.0$. This leads us to
explore alternative approaches to modeling the nonlinear power
spectrum on smaller scales. Before moving on, we note that there are
interesting new analytic approaches to solving the pertubation theory
at higher orders being developed, i.e. renormalized perturbation
theory and renormalization group theory
\citep{CrocceScoccimarro2006a,Bartoloetal2010,Bernardeauetal2010}.
These will push our analytic understanding deeper into the nonlinear
regime $k>0.2\kMpc$. However, we point out that the most recent
calculations in the literature, show only a very modest improvement
over the standard PT that we considered in this paper
\citep{Bartoloetal2010}. Moreover, whilst certainly interesting, these
techniques are unlikely to be able to probe the range of wavenumbers
required for future weak lensing missions ($k\in[0.01,100.0]\kMpc$).


\subsection{The halo model approach}\label{sec:halomodel}

The halo model was developed by a number of authors
\citep{Seljak2000,PeacockSmith2000,MaFry2000a}, and for a review see
\citep{CooraySheth2002}. In this model all of the mass in the Universe
is distributed into dark matter haloes, each labeled with some
physical properties. Typically one simply labels each halo by its
mass, however more complicated approaches can take into account, for
instance, halo shape \citep{SmithWatts2005}.

In the halo model the density field of dark matter may be written as a
sum over all haloes,
\be \rho(\bx)= \sum_{j}^{N}M_j U_j(\bx-\bx_j|M_j)\ ,\ee
where $N$ is the total number of haloes, $M_j$ and $\bx_j$ are the
mass and center of mass of the $j$th halo and
$U_j\equiv\rho_j(\bx_j)/M_j$ is the mass normalized density profile.
The statistics of the density field may be computed directly. In
particular for the density power spectrum it can be shown that it can
be written as the sum of two terms \citep[for more details
see][]{ScherrerBertschinger1991,SmithWatts2005}:
\be P(k) = P_{\1H}(k)+P_{\2H}(k)\ ,\label{eq:PowHaloModel}\ee
where the first term is referred to as the `1-Halo' term, which
describes the intra-clustering of dark matter particles within single
haloes; the second term is referred to as the `2-Halo' term, and
describes the clustering of particles in distinct haloes. They have
the explicit forms:
\ba P_{\1H}(k) & = & \frac{1}{\rhob^2}\int_{0}^{\infty} dM n(M) M^2
\left| U(k|M)\right|^2
\label{eq:Pow1H}\ ;\\
P_{\2H}(k) & = & \frac{1}{\rhob^2}\int_{0}^{\infty} \prod_{l=1}^{2}
\left\{dM_l n(M_l) M_l U_l(k|M_l)\right\}\nn \\ & & \times
\ P^{\hc\hc}_{\cent}(k|M_1,M_2)\label{eq:Pow2H} \ ,\ea
where the essential new ingredient is the power spectrum of halo
centers with masses $M_1$ and $M_2$, denoted
$P^{\hc\hc}_{\cent}(k|M_1,M_2)$.


\subsection{The halo-center power spectrum}

The power spectrum of halo centers contains all of the information for
the inter-clustering of haloes; precise knowledge of this term is
required to make accurate predictions on both large and small
scales. In principle, $P^{\hc\hc}_{\cent}(k|M_1,M_2)$ is a complicated
scale-dependent function of $M_1$, $M_2$ and $k$
\citep{Smithetal2007}.  The usual starting point for modeling this is
to assume the local deterministic biasing approach
\citep{FryGaztanaga1993,Coles1993}. This can be summarized as
follows. Consider the density field of all haloes with masses in the
range $M$ to $M+dM$, smoothed with some filter of scale $R$. We now
assume that this field can be related to the underlying dark matter
field, smoothed with the same filter, through some deterministic
mapping and that this mapping should apply independently of the
precise position, $\bx$, in the field: 
\be \delta^{\hc}(\bx|R,M)={\mathcal F}_{\left\{M,R\right\}}
\left[\delta(\bx|R)\right]\ ,\ee
where the sub-scripts on the function ${\mathcal F}$ indicate that it
depends on the mass of the haloes considered and the chosen filter
scale.  The filtered density field is
\be \delta(\bx|R)=\frac{1}{V}\int \dy\, \delta(\by)W(|\bx-\by|,R) \ ,\ee
$W(|x|,R)$ being some normalized filter.  Taylor expanding $\funk$
about the point $\delta=0$ yields
\be \funk\left[\delta(\bx|R)\right]=\sum_i 
\frac{b_i(M|R)}{i!}[\delta(\bx|R)]^i\ .\ee
We now assume that there is a certain filter scale above which $\funk$
is independent of both the scale considered and also the exact shape
of the filter function. Hence,
\be \delta^{\hc}(x|R,M)=\sum_{i=0}^{\infty}\frac{b_i(M)}{i!}
\left[\delta(\bx|R)\right]^{i}\ ,\label{eq:HaloDenFG}\ee
where the bias coefficients are 
\be b_i(M)\equiv\left.\frac{\partial^i\mathcal{F}_{\{M\}}
[\chi]}{\partial\chi^i}
\right|_{\chi=0} \ .\label{eq:biascoeffdeff}\ee

The bias coefficients from the Taylor series are not independent, but
obey the constraint, $\left<\delta^{\hc}(\bx|R)\right>=0$, which leads
to
\be b_0=-\frac{b_2}{2}\left<\delta^2\right>-\frac{b_3}{3!}
\left<\delta^3\right>-\dots-\frac{b_n}{n!}\left<\delta^n\right> \ .\ee
Thus, in general $b_0$ is non-vanishing and depends on the hierarchy
of moments. This allows us to re-write equation (\ref{eq:HaloDenFG})
as
\be \delta^{\hc}(x|R,M)=\sum_{i=1}^{\infty}\frac{b_i(M)}{i!}
\left\{\left[\delta(\bx|R)\right]^{i}-\left<\delta^{i}(\bx|R)\right>\right\}\
\label{eq:HaloDenFG2}\ .\ee
Nevertheless, we may remove $b_0$ from further consideration by
transforming to the Fourier domain, where it only contributes to
$\delta(\bk=0)$.

Thus the halo centre correlation function has the form:
\ba
\xi_{\hh,R}^{\rm cent}(r) & \equiv & \left<\delta^{\h}(\bx|R,M)\delta^{\h}(\bx+\br|R,M)\right>\nn \\
& = & b_1(M_1)b_1(M_2)\xi_R(r) +\frac{1}{6}\left[b_1(M_1)b_3(M_2)\right.
\nn \\ 
&  & \hspace{-0.7cm} \left. \frac{}{}+b_3(M_1)b_1(M_2)\right]
\left<\delta_R(\bx)\delta^3_R(\bx+\br)\right> \nn \\ 
&  & \hspace{-0.5cm} 
+\frac{b_2(M_1)b_2(M_2)}{4}\left<\delta^2_R(\bx)\delta^2_R(\bx+\br)\right>+ 
\dots \ .\label{eq:Lin2HaloXi} 
\ea
where $\xi_R(r)$ is the nonlinear matter correlation function smoothed
on scale $R$. Fourier transforming the above expression we obtain the
halo center power spectrum 
\citep{FryGaztanaga1993,MoWhite1996,Smithetal2007}:
\be P^{\rm \hc\hc}_{\cent}(k|M_1,M_2)=b_1(M_1)b_1(M_2)P_{\rm NL}(k|R)+
    {\mathcal O}(b_2,\dots) \ ,\label{eq:Lin2Halo} \ee
where the parameters $b_i$ are the nonlinear bias coefficients and
$P_{\rm NL}(k|R)$ is the nonlinear matter power spectrum smoothed on
scale $R$. It has recently been proposed that, for PNG, the halo bias
is also a function of the local gravitational potential
\citep{McDonald2008,GiannantonioPorciani2010}, we shall not explore
this possibility here, but simply note that it should give rise to the
same scale dependence of the linear bias. Further, since this is a
first order attempt to calculate the effects of PNG on the matter
clustering in the halo model, we shall restrict our attention to the
case of linear bias and so neglect terms $b_i$ with $i>1$. Whereupon,
$P^{\rm \hc\hc}_{\cent}$ becomes a separable function of mass and
scale. We present details of the $b_1(M,\FNL)$ model in
\S\ref{sec:ingred}. For $P_{\rm NL}(k,R,\FNL)$, we make the simple
approximation:
\be P_{\rm NL}(k|R,\FNL)=W^2(kR) P_{\tt halofit}(k) \beta_{\rm
  PT}(k,\FNL) \ee
where $W(kR)$ is a smoothing function, $P_{\tt halofit}(k)$ is the
nonlinear matter power spectrum model of \citep{Smithetal2003}, valid
for Gaussian initial conditions, and where $\beta_{\rm PT}$ was defined
earlier in \Eqn{eq:beta}.

As was argued in \citep{TakadaJain2003,Tinkeretal2005,Smithetal2007}
another essential component of the inter-clustering of haloes is halo
exclusion. That is, one must remove the correlations which arise on
scales inside the sum of the virial radii of the two haloes $M_1$ and
$M_2$. As was shown in \citep{Smithetal2007}, this effect can formally
be written:
\be \xi^{\hc\hc}_{\cent}(r|M_1,M_2)=-1 \ ; \ \
(r<r_{\vir,1}+r_{\vir,2})\ , \label{eq:exclusion}\ee
where $r_{\vir}$ is the virial radius of a halo and where
$\xi^{\hc\hc}$ is the correlation function of dark matter halo
centres, defined:
$\xi^{\hc\hc}_{\cent}(r|M_1,M_2)\equiv\left<\delta^{\h}(\bx|M_1)\delta^{\h}(\bx+\br|M_2)\right>$. The
$-1$ in the above is simply the value that $\xi$ must obtain in order
for the joint probability of finding halo center separations $r<r_{\rm
  vir,1}+r_{\rm vir,2}$ to be zero. In the literature, various
approximate schemes have been proposed to model the exclusion effect
\citep{TakadaJain2003,Tinkeretal2005}, these involve placing a cut-off
in the upper limit of the mass integrals in the 2-Halo term. We shall
not follow such schemes, since these approaches do not reproduce the
correct power spectrum asymptotics for the exact calculation, which we
show below. Instead we follow \citep{Smithetal2007}, and evaluate the
above expression exactly. In this case the halo center power spectrum
can be written in terms of the correlation function of halo centres
as:
\be
P^{\hc\hc}_{\cent}(k|R)  =  \int \dr \xi^{\hc\hc}_{\cent}(k|M_1,M_2,R) j_0(kr) 
\ee
On inserting \Eqn{eq:exclusion} for scales inside
$r_{\vir,1}+r_{\vir,2}$ and the relation
$\xi^{\hc\hc}_{\cent}(k|M_1,M_2,R)=b(M_1)b(M_2)\xi(r|R)$ on larger
scales, where $\xi(r|R)$ is the dark matter correlation function
smoothed on the scale $R$, then we find
\ba P^{\hc\hc}_{\cent}(k|R) & = &  
\int_{r_{\vir,1}+r_{\vir,2}}^{\infty} \dr b(M_1)b(M_2)\xi(r|R) j_0(kr)\nn \\
&  & 
\hspace{-1.5cm} + \int_0^{r_{\vir,1}+r_{\vir,2}} \dr (-1) j_0(kr) 
\nn\\ 
&  & \hspace{-1.5cm} =
\int_{0}^{\infty} \dr b(M_1)b(M_2)\xi(r|R) j_0(kr) \nn \\
&  & \hspace{-1.5cm} - \int_0^{r_{\vir,1}+r_{\vir,2}} \dr 
\left[1+b(M_1)b(M_2)\xi(r|R)\right] j_0(kr) \nn\\ 
&  &  \hspace{-1.5cm} =
P^{\rm NoExc,\hc\hc}_{\cent}(k)  - P^{\rm Exc,\hc\hc}_{\cent}(k)  \ .
\ea
The first term in the last line of the above equation represents the
usual expression for the clustering of halo centers, and the second
term represents the correction due to halo exclusion:
\ba 
\!\!\!\!P^{\rm NoExc,\hc\hc}_{\cent} & \equiv & b(M_1)b(M_2)P_{\Lin}(k) \ ; \\
\!\!\!\!P^{\rm Exc,\hc\hc}_{\cent}  & \equiv &
\int_0^{r_{\vir,1}+r_{\vir,2}} \hspace{-1cm}\dr \left[1+b(M_1)b(M_2)\xi(r)\right] j_0(kr) \ .
\ea
Taking $s\equiv s(M_1,M_2)\equiv r_{\vir,1}+r_{\vir,2}$ and $y\equiv ks$, we
may deduce the following asymptotic properties for $P^{\rm
  Exc,\hc\hc}_{\cent}$:\\

\noindent {$\bullet$ \it Large-scale limit}: for the case
$k\rightarrow0$, we have that $j_0(kr)\rightarrow 1$ and so
\ba
& & \lim_{y\rightarrow0} P^{\rm exc,\hc\hc}_{\cent}(k|M_1,M_2) \nn \\
& & \hspace{1cm} = \int_0^{s} \dr \left[1+b(M_1)b(M_2)\xi(r)\right]\nn \\
& & \hspace{1cm} =  V(s)\left[1+b(M_1)b(M_2)\overline{\xi}(s)\right]\label{eq:2HExcLS} ,
\ea
where $V(s)=4\pi s^3/3$ is the volume of the exclusion sphere for the
haloes $M_1$ and $M_2$ and where $\overline{\xi}$ is the volume
averaged correlation function. This appears as white noise power
contribution, and so it acts to reduce any large-scale shot noise
component.\\

\noindent {$\bullet$ \it Small-scale limit}: for the case
$k\rightarrow\infty$, we have that
\ba 
& & \lim_{k\rightarrow\infty} P^{\rm exc,\hc\hc}_{\cent}(k|M_1,M_2) \nn \\
& & = \lim_{k\rightarrow\infty} 
\int_0^{s} \dr \left[1+b(M_1)b(M_2)\xi(r) \right]j_0(kr)
\nn \\
& & = \lim_{k\rightarrow\infty} \left\{ \frac{1}{k^3}\int_0^{ks} \dy j_0(y)\right\} \nn \\
& & \ \ \ +\lim_{k\rightarrow\infty}\left\{\frac{1}{k^3}\int_0^{ks} \dy' 
b(M_1)b(M_2)\xi(y'/k)j_0(y')\right\}
\nn \\
& &  =  (2\pi)^3\delta^{D}(\bk) + P^{\rm NoExc,\hc\hc}_{\cent}(k) \ .
\ea
Thus the effect of halo exclusion on small scales, is to exactly null
the 2-Halo term without exclusion.  \\

This leads us to write the full 2-Halo term as
\mbox{$P_{\2H}(\bk)-P_{\2H}^{\rm exc}(\bk)$}, where
\ba
\!\!\!P_{\2H}^{\rm exc}(k) & = & \frac{1}{\rhob^2}\int_{0}^{\infty} \prod_{l=1}^{2}
\left\{dM_l n(M_l) M_l U_l(\bk|M_l)\right\} \nn\\
&  &  \hspace{-1.0cm} \times \int_0^{r_{\vir,1}+r_{\vir,2}} \hspace{-0.3cm}  \dr 
\left[1+b(M_1)b(M_2)\xi(r)\right] j_0(kr) .
\ea

As a short aside, in Appendix \ref{sec:RPTHM} we forward the idea
that halo exclusion may resolve the well known problem of excess
large-scale power in the standard formulation of the halo model and
that after taking this into account the theory is consistent with
perturbation theory results, like RPT \cite{CrocceScoccimarro2008}.


\section{Halo model ingredients in non-Gaussian models}
\label{sec:ingred}

The model that we described in \S\ref{sec:halomodel}, specified
nothing about the cosmological model, other than that the end state of
gravitational clustering leads to the formation of a distribution of
haloes with some characteristic spectrum of masses, density profiles,
and that halo centers are clustered. Thus no extension of the
formalism is necessary in order to use the halo model to describe
clustering in more exotic models, such as PNG. However, what must
necessarily change are the ingredients of the model: the mass
function, the halo bias and the density profiles.  We now study these
in detail in the context of PNG.


\begin{figure*}
\centering{
  \includegraphics[width=7.5cm,clip=]{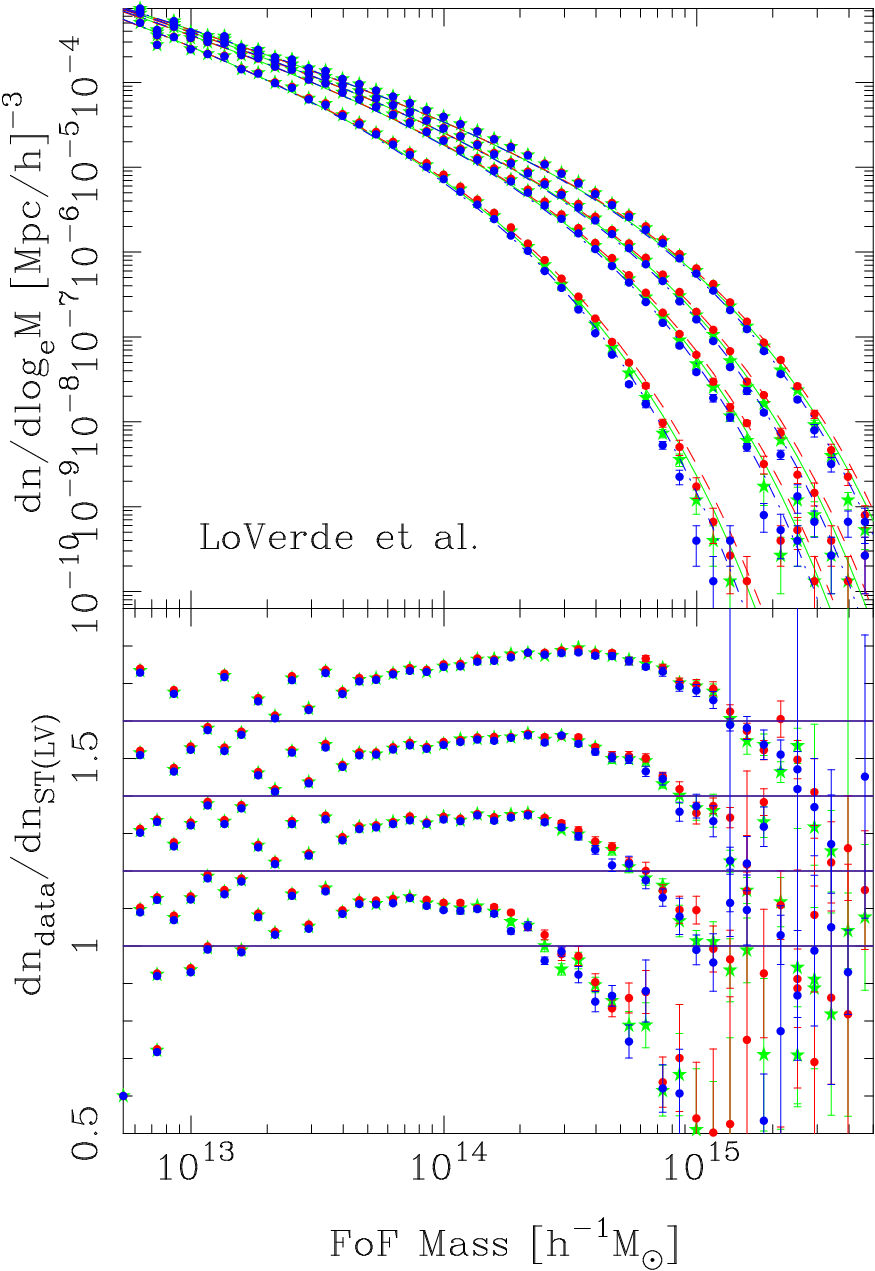}\hspace{0.2cm}
  \includegraphics[width=7.5cm,clip=]{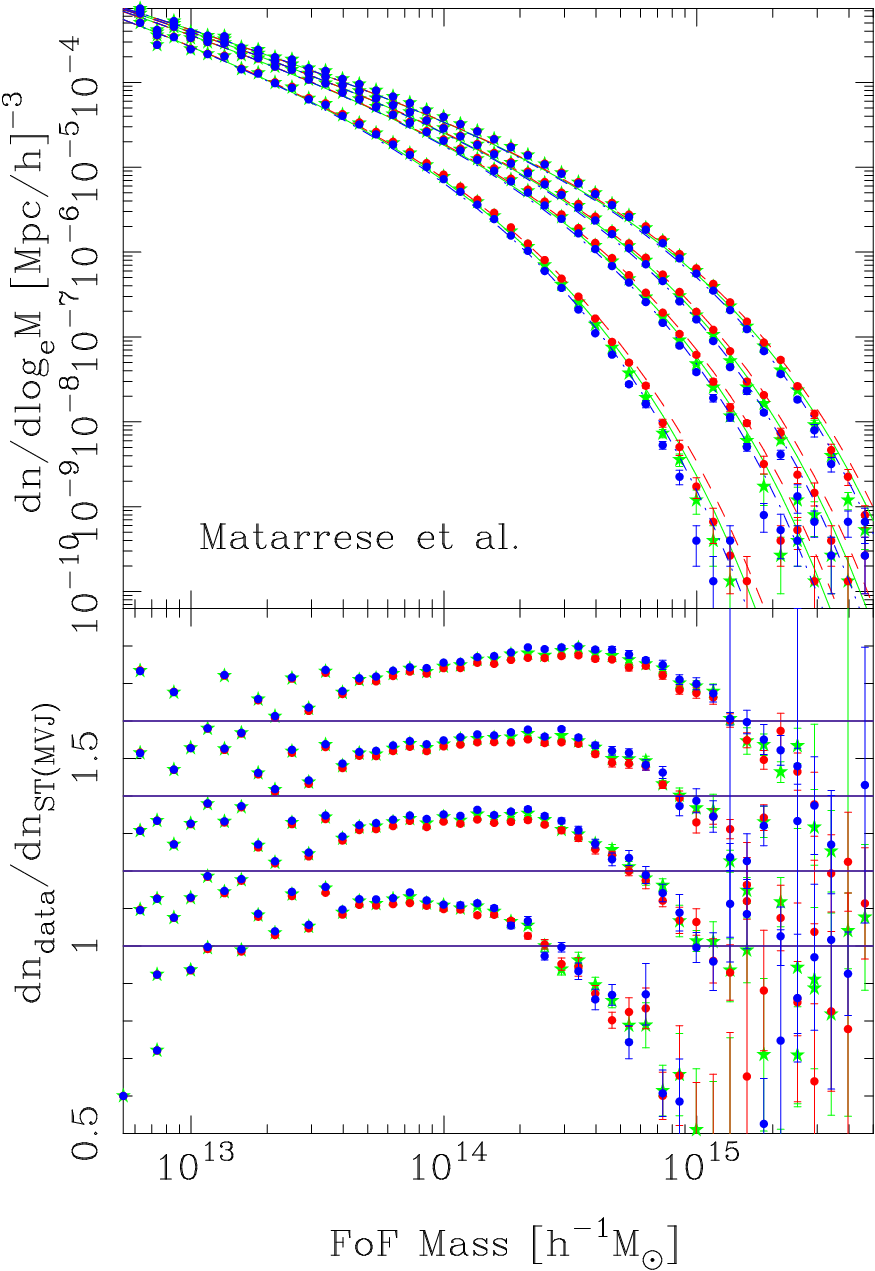}}
\caption{\small{Mass functions of haloes in models with local
    primordial non-Gaussianity as a function of FoF ($b=0.2$) halo
    mass and at several redshifts. Left and right panels compare
    estimates from the ensemble of $N$-body simulations with the
    theoretical predictions of \citet{LoVerdeetal2008} and
    \citet{Matarreseetal2000}, respectively. Gaussian mass function
    predictions are given by \citet{ShethTormen1999}. {\it Top panels}
    show the absolute mass function. The green, red and blue points
    with errors denote estimates for $\FNL=\{0,+100,-100\}$. The
    solid, dash and dot dash lines represent the predictions as given
    by \Eqn{eq:massfunNG} for $\FNL=\{0,100,-100\}$. From bottom to
    top the points show results for redshifts
    $z\in\{1.0,\,0.5,\,0.28,\,0.0\}$.  {\it Bottom panels:} ratio of
    the estimated mass functions with their respective theoretical
    predictions. For clarity, we have offset the results for redshifts
    $z\in\{1.0,\, 0.5,\, 0.28,\, 0.0\}$ by $\{0.0,\, 0.2,\, 0.4,\,
    0.6\}$ in the positive $y$-direction.}\label{fig:MF}}
\vspace{0.5cm} \centering{
  \includegraphics[width=7.5cm,clip=]{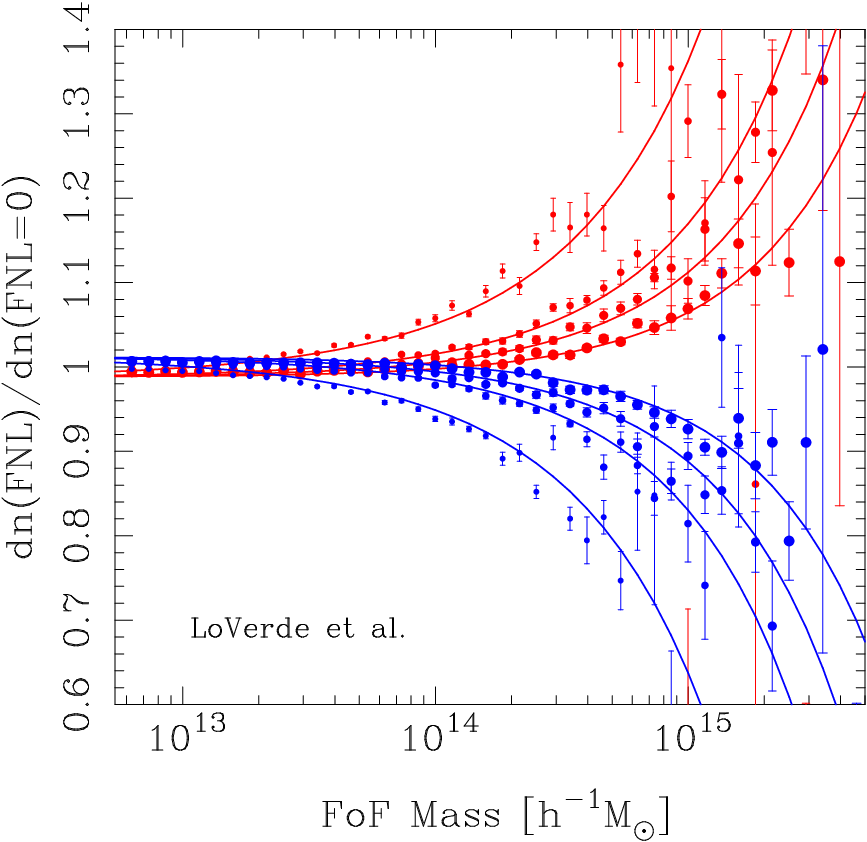}\hspace{0.2cm}
  \includegraphics[width=7.5cm,clip=]{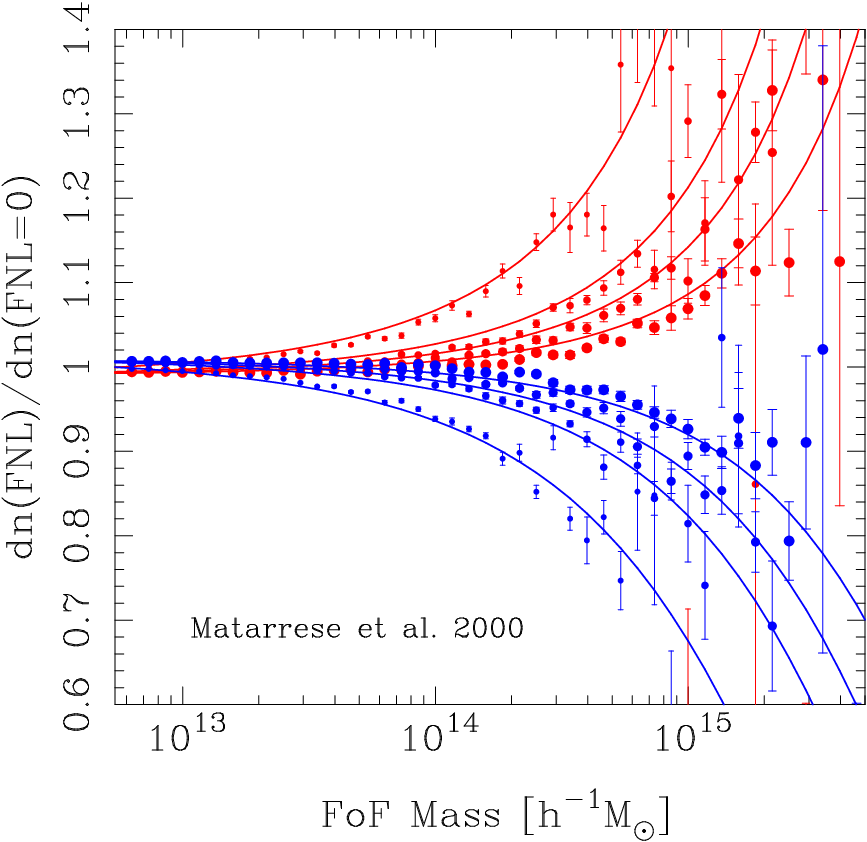}}
\caption{\small{Ratio of the non-Gaussian to Gaussian mass functions
    as a function of FoF ($b=0.2$) halo mass. Left and right panels
    show compare the simulations with the theoretical models of
    \citep{LoVerdeetal2008} and \citep{Matarreseetal2000},
    respectively. The points with errors denote the ensemble averages
    of the mass function ratios measured at expansion factors
    $z\in\left[1.0,0.5,0.28,0.0\right]$, where larger point symbols
    denote later times. Solid lines denote theoretical models. The
    theory predictions were generated using:
    $\delta_c\rightarrow\sqrt{q}\delta_c$.}\label{fig:MFRatio}}
\end{figure*}


\subsection{The halo mass function}

The mass function of dark matter haloes in models of structure
formation from Gaussian initial conditions has been widely studied
over the past decades
\cite{PressSchechter1974,Bondetal1991,ShethTormen1999,Warrenetal2006,
  Crocceetal2010,Maneraetal2010}. Conventionally, the mass function is
represented:
\be \frac{dn}{d\log M}=\frac{\rhob}{M} f(\nu) \frac{d\log\nu}{d \log M}\ ; 
\hspace{1cm} \nu\equiv \frac{\delta_c(z)}{\sigma(M)}\ ,\ee
where for the Press-Schechter (PS) and Sheth \& Tormen (ST) mass
function we have:
\ba 
f_{\rm PS}(\nu)  & = & 
      \sqrt{\frac{2}{\pi}}\nu \exp\left[-\frac{\nu^2}{2}\right]\ ;\\ 
f_{\rm ST}(\nu)  & = & 
      A \sqrt{\frac{2}{\pi}} \sqrt{q}\nu \left[1+(\sqrt{q}\nu)^{-2p}\right]
\exp\left[-\frac{q\nu^2}{2}\right] \label{eq:STMF}\ .
\ea
For the ST mass function the amplitude parameter $A$ is determined
from the constraint $\int d\log \nu f(\nu) =1$: which leads to
$A^{-1}=\left\{1+2^{-p}\Gamma\left[0.5-p\right]/\Gamma\left[0.5\right]\right\}$.
ST's original parameters are: $\{A=0.3222, p=0.3, q=0.707\}$. The
derivative term on the right-hand-side can be calculated from
\Eqn{eq:sigdiff} in Appendix~\ref{app:skew}. \Fig{fig:skewness} shows
the derivative of \Eqn{eq:sigdiff} computed for the WMAP5 parameters
compared to the skewness due to PNG.

A number of authors have studied the effects of PNG of the local type
on the mass function of dark matter haloes
\citep{Matarreseetal2000,LoVerdeetal2008,Dalaletal2008,Grossietal2009,LamSheth2009,Pillepichetal2010,MaggioreRiotto2010}.
The most important task in extending the Press-Schechter framework is
to find an analytic expression for the 1-point probability density
function (PDF) of the smoothed matter fluctuations.
\citep[][hereafter MVJ]{Matarreseetal2000} gave the first formal
derivation using a path-integral approach.  \citep[][hereafter
LV]{LoVerdeetal2008} used the Edgeworth expansion (or more simply the
Gram-Charlier Type Ia series) to recover the PDF. The key idea of
these expansions is to write the characteristic function of the
non-Gaussian PDF to be approximated in terms of the characteristic
function of the Gaussian PDF, and to then recover the non-Gaussian PDF
through the inverse Fourier transform method
\citep[see][]{BlinnikovMoessner1998}.

For small amounts of PNG the mass function can be written:
\be \frac{dn(M\FNL)}{d\log M}= \frac{dn_{\rm ST}(M)}{d\log M} R(\nu,\FNL)\
\label{eq:massfunNG},\ee
where the ratio of the non-Gaussian to the Gaussian mass functions of
MVJ and LV can be written:
\begin{widetext}
\ba 
R_{\rm MVJ}\left[\nu,\FNL\right] & = &
\exp
\left[\frac{\tilde{\delta}_{\rm c}^3(a) S_3(M,a_0)}{6\sigma^2(M,a_0)}\right]
\left|
\frac{1}{6}\frac{\tilde{\delta}_{\rm c}(a)}{\sqrt{1-\tilde{\delta}_{\rm c}(a)S_3(M,a_0)/3}}
\frac{dS_3(M,a_0)}{d\log\sigma}+\sqrt{1-\frac{1}{3}\tilde{\delta}_{\rm c}(a)S_3(M,a_0)}\right| \\
R_{\rm LV}\left[\nu,\FNL\right] & = &
1+\frac{1}{6}\sigma(M|a_0)S_3(M|a_0)\left[\tilde{\nu}^3(a)-3\tilde{\nu}(a)\right]
+\frac{1}{6}\frac{d\left[\sigma(M|a_0)S_3(M|a_0)\right]}{d \log\sigma}
\left[\tilde{\nu}(a)-\frac{1}{\tilde{\nu}(a)}\right]
\ea
\end{widetext}
where in the above equations $\tilde{\nu}\equiv\tilde{\delta}_{\rm
  c}(a)/\sigma(M,a_0)$, where we use the rescaled linear collapse
density: $\tilde{\delta}_{\rm c}(a)=\sqrt{q}\delta_{\rm c}/D(a)$. In
the spherical collapse model, the linearly extrapolated density
threshold for collapse is $\delta_{\rm c}=1.686$. Note that, as in the
Press-Schechter formalism, we have chosen to evaluate the variance and
the skewness of the density field at the present time $a_0$, and in so
doing have transferred the time dependence of the theory to the
collapse barrier $\delta_{\rm c}(a)=\delta_{\rm c}(a_0)/D(a)$.

Note that these formula are not the same as those originally derived
by MVJ and LV, but differ by the $\sqrt{q}$ in the definition of the
peak height. This was introduced by \citet{Grossietal2009} in a
heuristic way in order to obtain a good fit to the FoF mass functions
in their numerical simulations. They conjecture that it appears for
the same reason as it appears in the ST mass function \citep[see
also][]{LamSheth2009}. The true orginis for such a factor are unclear,
since for the case haloes identified through a bound spherical
over-density criterion no $q$ correction is required
\citep{Desjacquesetal2009}.

Figure \ref{fig:MF} compares the mass function of FoF ($b=0.2$) dark
matter haloes measured from the ensemble of simulations at redshifts:
$z\in\left[1.0,0.5,0.28,0.0\right]$ . The left and right panels
compare the simulation estimates with the predictions from the LV and
MVJ models, respectively. The top panels show the absolute mass
function and bottom panels show the ratio with respect to the
theory. In these plots we take the Gaussian model for the mass
function to be that as given by \citep{ShethTormen1999}. Note that the
errors on the points show the error on the mean, i.e. the box-to-box
varaince divided by the square root of the number of realizations:
$\sigma/\sqrt{N_{\rm ensemb}}\sim\sigma/3.5$.

A number of important points may be noted. Firstly, none of these
models fit the data well \citep{GiannantonioPorciani2010}, however
this can be mainly attributed to the fact that the ST mass function
does not fit well the Gaussian simulation data: we see a
$\sim$10--20\% excess in the number of intermediate mass haloes
$M_{*}<M<100M_{*}$ and strong suppression in the numbers of high mass
haloes $M>100M_{*}$. Secondly, we note that as expected, the model
with $\FNL>0$ $(\FNL<0)$ produces an excess (reduction) in the number
of high mass haloes relative to the Gaussian case. The predictions
capture these trends. However, as can be seen from the lower section
of the left figure, the model of LV produces the same locus of points
for all of the models $\FNL\in\left[\pm100,0\right]$ and for all
redshifts. Whereas that of MVJ, being almost equally as good, produces
a slightly different offset for each $\FNL$ model and at different
redshifts. This leads us to conclude that in order to accurately
predict the mass function in non-Gaussian models over a wide range of
masses and redshifts, one simply requires an accurate fit to the
Gaussian model, combined with the ratio model of LV.

Figure \ref{fig:MFRatio} further emphasizes this point. Here we show
the fractional mass function for the FoF dark matter haloes measured
in the simulations at the four expansion factors. For each simulation
we compute the mass function of haloes and take the ratio of this
estimate with respect to the Gaussian model. These results are
averaged over the 12 realizations and the error are computed on the
mean. On comparing these estimates with the theoretical predictions
from the models of LV and MVJ we find excellent agreement when the
shift parameter $\delta_c\rightarrow\sqrt{q}\delta_c$ is used, as
advocated in \citep{Grossietal2009}. There is a small preference to
the model of LV, especially at late times for high mass haloes in the
model with $\FNL=100$. Coupled with the fact that the mass function
ratios with respect to the ST+LV model are well behaved, we shall
hereafter adopt the LV model for all our halo model calculations. 

Finally, we note that the ST mass function is preferred over other
commonly used expressions such as that of \citep{Warrenetal2006},
since it obeys the important property that when integrated over all
masses, one recovers the mean matter density.


\begin{figure*}
\centering{
  \includegraphics[width=7.8cm,clip=]{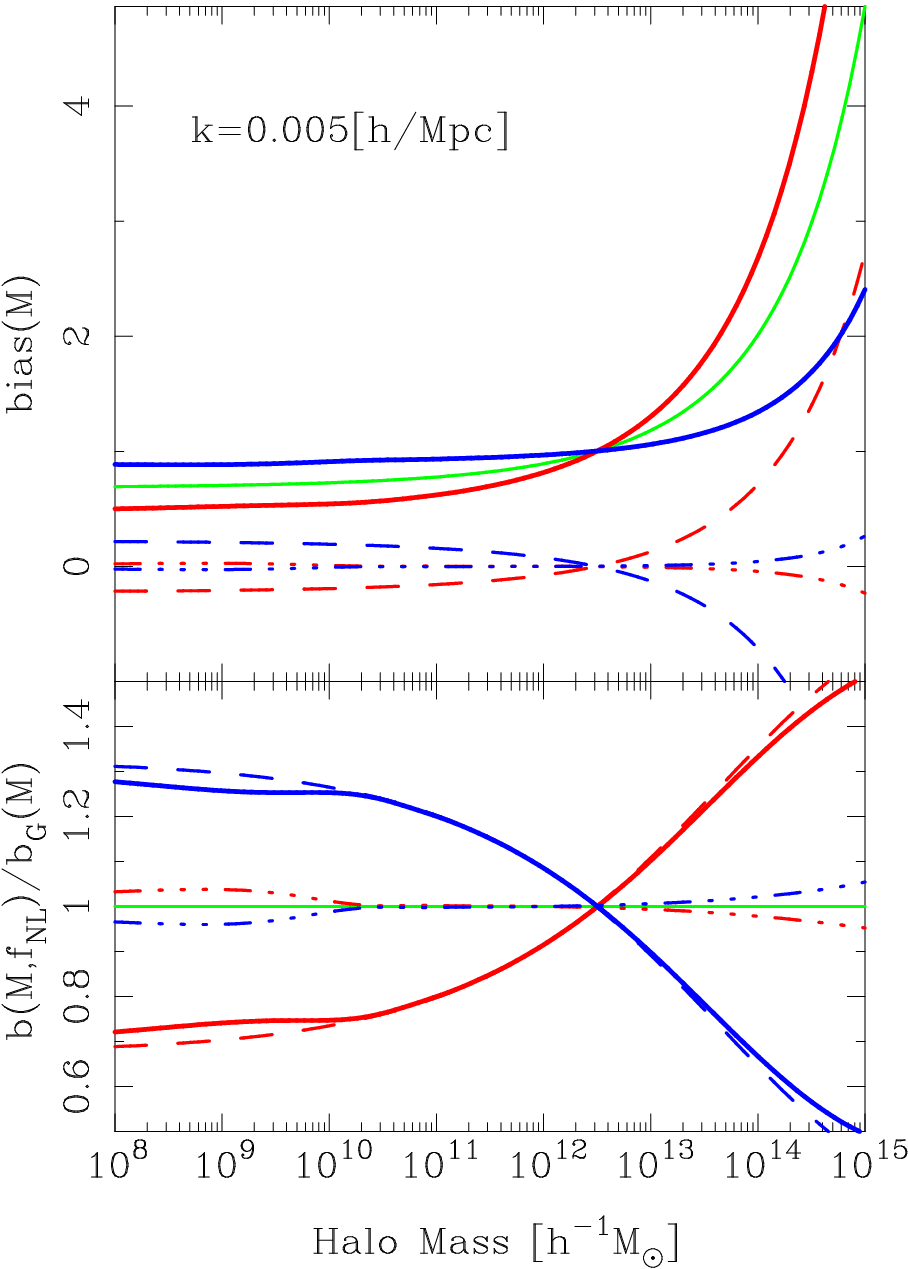}\hspace{0.3cm}
  \includegraphics[width=7.8cm,clip=]{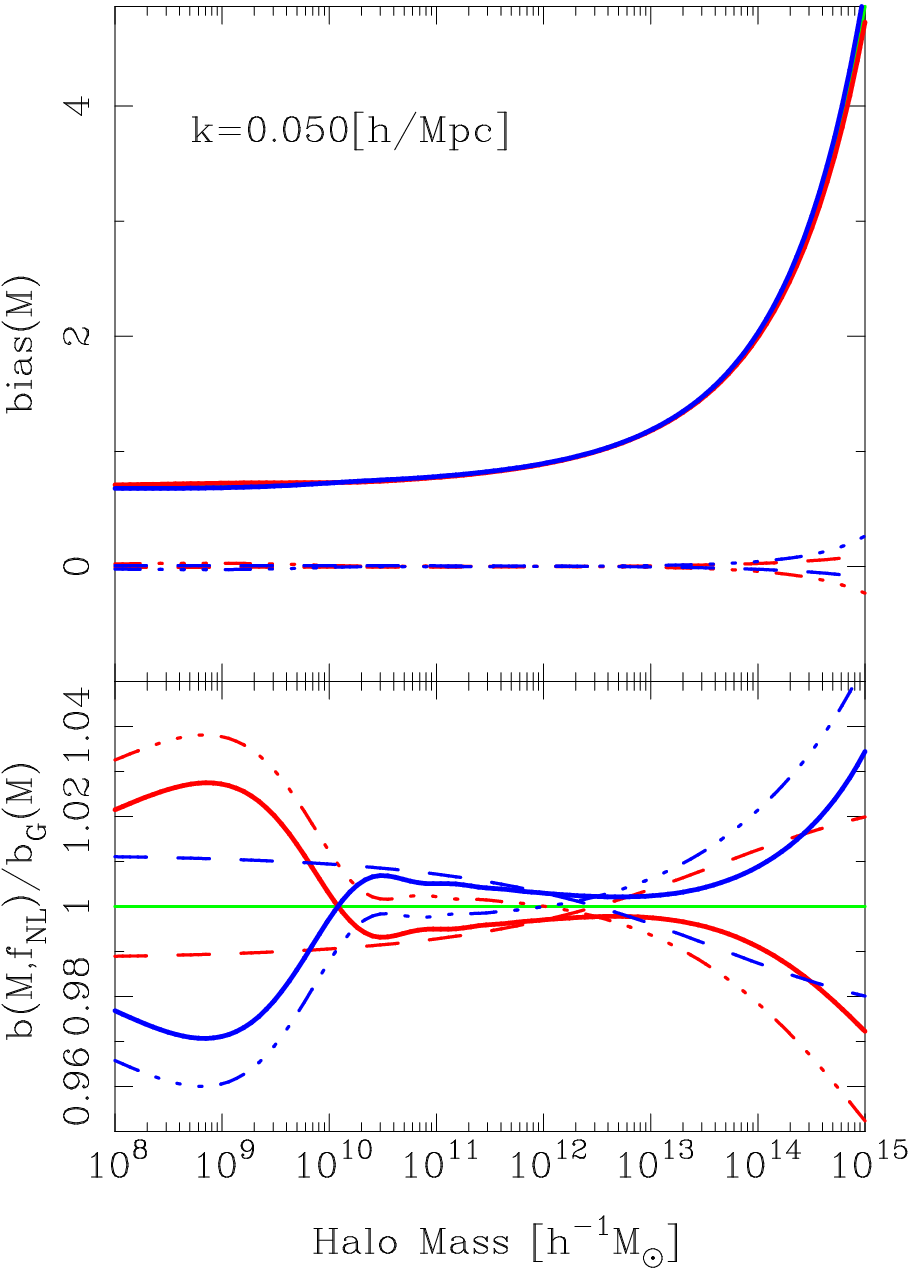}}
\caption{\small{Mass dependence of halo bias in models with primordial
    non-Gaussianity. The left and right panels, show results for the
    halo bias on scales $k=\{0.005,0.05\}\kMpc$, respectively. {\it
      Top panels:} absolute bias vs halo mass. The green, red and blue
    lines correspond to the cases $\FNL=\{0,100,-100\}$ respectively;
    dashed lines denote the scale-dependent contribution $\Delta
    b_{\kappa}(k,\FNL)$; and triple-dot-dash lines denote the
    scale-independent contribution $\Delta b_{\rm I}$. {\it Bottom
      panels:} show the ratio of the non-Gaussian bias to the Gaussian
    bias. Solid lines denote the bias ratio for the sum of all
    components; dashed lines denote the same but excluding the
    scale-independent bias; triple-dot-dash lines denote the same but
    excluding the scale-dependent bias.}\label{fig:BiasMass}}
\end{figure*}


\subsection{Halo bias}

The halo model also requires us to specify how the centers of dark
matter haloes of different masses cluster with respect to each other.
As described by \Eqn{eq:Lin2Halo}, for the Gaussian model, and at
first order in the dark matter density, halo and matter density
perturbations can be related through a scale-independent bias factor
$b(M)$.  Following \citep{MoWhite1996,ShethTormen1999}, an application
of the peak-background split approximation enables one to calculate
$b(M)$ from a given mass function. For the ST mass function the
Gaussian bias has the form:
\be b_{\rm ST}(\nu)= 1 + \frac{q\nu^2
  -1}{\delta_c(z)}+\frac{2p}{1+(q\nu^2)^p} \ ;
\hspace{0.4cm}\nu\equiv\frac{\delta_c(z)}{\sigma(M,z=0)}\ ,\ee
where the parameters $\{p,q\}$ are as in \Eqn{eq:STMF}.

As was summarized in \citep{Desjacquesetal2009}, with PNG there are
two main effects on the bias. Firstly, local non-Gaussianity induces a
scale-dependent correction factor on extremely large scales
$k<0.02\kMpc$
\citep{Dalaletal2008,MatarreseVerde2008,GiannantonioPorciani2010}.
Secondly, assuming that the peak background split holds, then there is
also a scale-independent correction to the bias. This arises due to
the fact that the mass function changes with $\FNL$ and consequently
the typical halo mass changes and hence the bias
\citep{Slosaretal2008,AfshordiTolley2008}.  Thus the non-Gaussian bias
may be written:
\be
b_{\rm NG}(k,M,\FNL)=b_{\rm G}(M)+ \Delta b_{\kappa}(k,M,\FNL) + \Delta b_{\rm I}(M,\FNL)
\ .\ee

The scale-dependent bias term $\Delta b_{\kappa}(k,M,\FNL)$ can be
written,
\be \Delta b_{\kappa}(k,M,a) \equiv 2\FNL \left[b_{\rm
    G}(M,a)-1\right] \frac{3 \Omega_m H_0^2 \delta_c(a)}
{2D(a)c^2T(k)k^2} \ .\label{eq:SDB}\ee
In the original derivation of \citep{Dalaletal2008}, as was pointed
out by \citet{MatarreseVerde2008}, a factor of the transfer function
was missing, and it has been added in the above
expression. Furthermore, in the original derivation one sees that the
term in brackets involves just the Gaussian bias $b_{\rm
  G}(M,a)$. However, as we will show in \Eqn{eq:SDB2}, this should in
fact be the sum of the Gaussian bias plus the scale-independent bias
correction due to PNG.

The scale-independent bias correction, $\Delta b_{\rm I}$, can be
written
\be
\Delta b_{\rm I}  =  -\frac{1}{\sigma}
\frac{\partial\ln[R(\nu,\FNL)]}{\partial\nu}  \ , \ee
where $R$ is the fractional correction to the mass function.  For the
case of \citet{LoVerdeetal2008} the scale-independent bias has the
form,
\ba \Delta b_{\rm I}^{\rm LV} & = & -\frac{1}{6 \sigma R_{\rm LV}
}\left[\frac{}{}- \frac{d^2(\sigma
S_3)}{d\ln\nu^2}\left(1-\frac{1}{\nu^2}\right) \right. \nn \\ & + &
\left.\frac{d(\sigma
S_3)}{d\ln\nu}\left(\nu^2-4-\frac{1}{\nu^2}\right) +3\,\sigma
S_3\left(\nu^2-1\right) \right] \label{eq:SIB} \ .  \ea
It is worth noticing that $\Delta b_{\rm I}(\FNL)$ has a sign opposite
to that of $\FNL$ (because the bias decreases when the mass function
goes up). 

The scale dependence of the above non-Gaussian bias model, for haloes
with masses above $M\sim10^{13}\Msol$, was recently tested against a
suite of high-resolution $N$-body simulations by
\citep{Desjacquesetal2009,Pillepichetal2010,Grossietal2009}. For the
cross-power spectrum of haloes and matter, it was shown to work at a
precision of better than $10\%$. However, the form of the bias has not
been investigated for halo masses substantially lower than $M*$. In
order to make a halo model calculation for the mass power spectrum one
is required to average over haloes of all masses and so it is
important to understand the behavior for $\nu<1$ as well as for
$\nu>1$. Note that whilst the mass function of \citep{LoVerdeetal2008}
should only be trusted for $\nu\gtrsim 1$ and hence also the bias
expansion, since we see no deviations at $\nu\sim1$ we shall assume
that these expansions can be trusted to lower $\nu$.

In \Fig{fig:BiasMass}, using \Eqns{eq:SIB}{eq:SDB}, we show how the
halo bias depends on halo mass, and we extrapolate these relations to
masses $M/M_*\lesssim1$ (note that in evaluating the scale-independent
bias we have made the approximation $d^2(\sigma
S_3)/d\ln\nu^2=0$. Since the bias is also scale-dependent, we show the
results for spatial scales $k\in\{0.005,0.05\}\kMpc$.

On very large scales $k\sim0.005\kMpc$ (left panel
\Fig{fig:BiasMass}), we see that the scale-dependent bias correction
$\Delta b_{\kappa}$ strongly modulates the Gaussian bias, increasing
(decreasing) the bias of high mass haloes for $\FNL>0$ $(\FNL<0)$.
For low mass haloes with $\nu<1$ this trend reverses. On smaller
scales $k\sim0.005\kMpc$ (right panel \Fig{fig:BiasMass}), the
scale-dependent bias has been strongly suppressed and the
scale-independent term starts to dominate. The consequence of this is
that high mass haloes in models with $\FNL>0$ ($\FNL<0$) have a
slightly lower (higher) bias than in the Gaussian case. Again this
trend is reversed for haloes with $\nu<1$.

In evaluating the 2-Halo term we must compute integrals of the type as
given in \Eqn{eq:Pow2H}. If we consider again the large-scale limit
\mbox{$U(k|M,\FNL)\rightarrow1$} as $k\rightarrow0$, then this places
the following conditions on the Gaussian and non-Gaussian mass
functions and bias:
\ba
& & \int dM M n_{\rm G}(M) b_{\rm G}(M)=\rhob \ ;\label{eq:cond1}\\
& & \int dM M n_{\rm NG}(M) b_{\rm NG}(M)=\rhob \ \label{eq:cond2}.
\ea
These relations arise simply from the fact that 
\ba 
\bar{n}(\bx,M) & = & \bar{n}(M)\left[1+\delta^{\rm h}(\bx|M)\right] \ .
\label{eq:haloden}\ea
Taking the non-linear local bias model of \citep{FryGaztanaga1993} for
the halo density field, $\delta^{\rm h}(\bx|M)=\sum_{i=0}^{\infty}
b_i(M)\delta^i(\bx)/i!$, and on integrating \Eqn{eq:haloden} over all
haloes, weighted by the mass, then we must recover the mass density
field:
\ba
\rhob\left[1+\delta(\bx)\right] & = & 
\int_{0}^{\infty} dM M \bar{n}(M) \left[1+\delta^{\rm h}(\bx|M)\right] \nn \\
& & \hspace{-1.5cm} = \int_{0}^{\infty} dM M \bar{n}(M)\left[1+\sum_{i=0}^{\infty} 
\frac{b_i(M)}{i!}\delta^i(\bx)\right] \ .
\ea
The integral resulting from the first term in the bracket on the right
side gives us the normalization condition for the mass function
$ \int_{0}^{\infty} dM M \bar{n}(M)=\rhob$,
and the integral of the second term gives:
\ba
\delta(\bx)=\frac{1}{\rhob} \int_{0}^{\infty} dM M \bar{n}(M)
\sum_{i=0}^{\infty} \frac{b_i(M)}{i!}\delta^i(\bx) \label{eq:cond3}\ .
\ea
This can only be true if and only if \Eqns{eq:cond1}{eq:cond2} are
true. As a caveat we also note the further condition:
\be \int dM M n(M) b_i(M)=0 \ \ (i\ne1)\ .\ee
Following \Eqn{eq:cond2}, for PNG this then implies the further two
conditions that:
\ba 
& & \int dM M n_{\rm NG}(M) \left[b_{\rm G}(M)+\Delta
  b_{I}(M|\FNL)\right]=\rhob\ .\\
& & \int dM M n_{\rm NG}(M) \Delta b_{\kappa}(k,M|\FNL) = 0 \label{eq:SDBiasZero}\ .
\ea
The last condition can only be correct if $b_{\rm G}$ in \Eqn{eq:SDB}
becomes $b_{\rm G}+\Delta b_I$, i.e. 
\be \Delta b_{\kappa}(k,M) \equiv \FNL \left[b_{\rm G}(M)+\Delta b_I(M)-1\right]
\frac{3 \Omega_m H_0^2 \delta_c} {D c^2T(k)k^2} \ .\label{eq:SDB2}\ee
This correction has not been pointed out in any previous work and we
expect it to affect the scale-dependence of the bias. We note that
this should also lead to a small asymmetry between the predictions for
$\FNL>0$ and those for $\FNL<0$.

Returning to the halo model, given \Eqn{eq:SDBiasZero} and the fact
that as $k$ becomes larger $\Delta b_{\kappa}$ becomes less important
(c.f. \Fig{fig:BiasMass}), we make the approximation for the dark
matter, that $\Delta b_{\kappa}=0$ on all scales. For an alternative
approach see to imposing \Eqn{eq:SDBiasZero}
\citep{FedeliMoscardini2010}.


\begin{figure*}
\centering{
  \includegraphics[width=5.1cm,clip=]{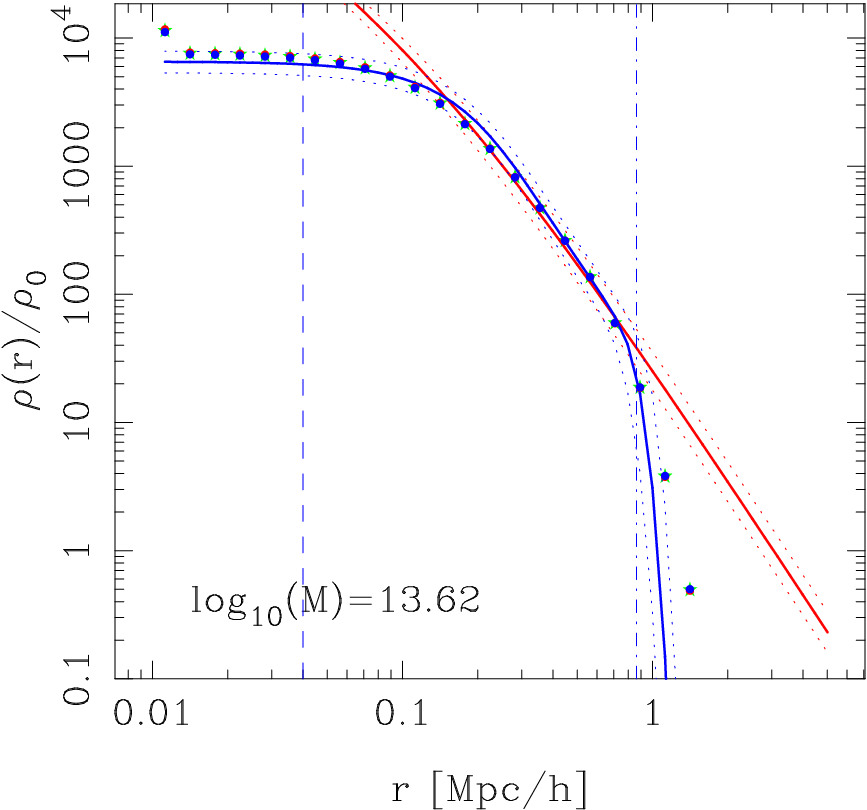}
  \includegraphics[width=5.1cm,clip=]{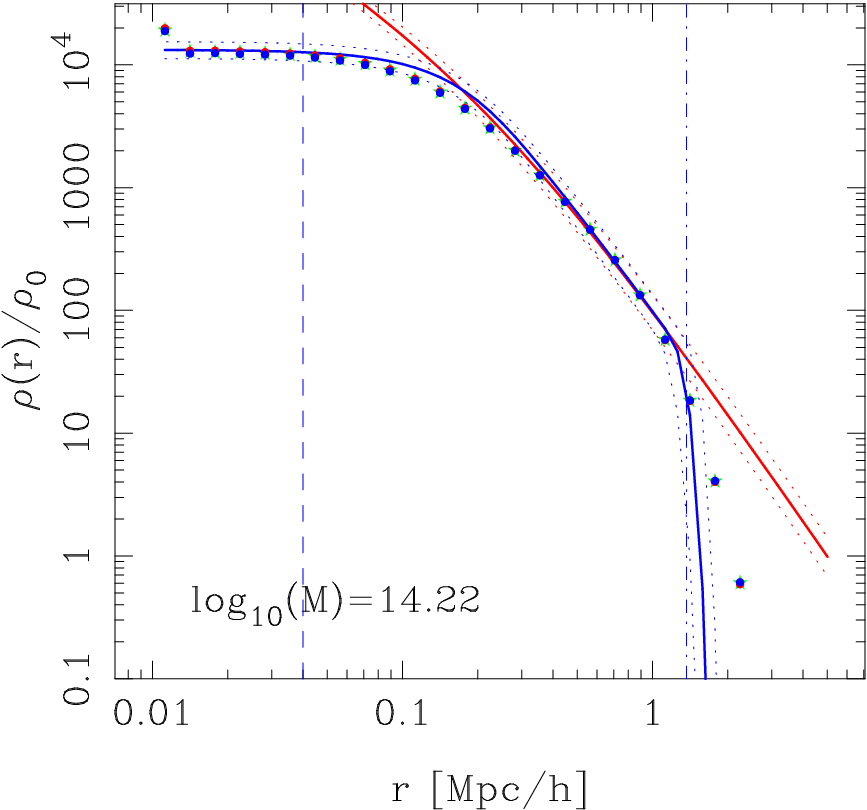}}
\\ \centering{
  \includegraphics[width=5.1cm,clip=]{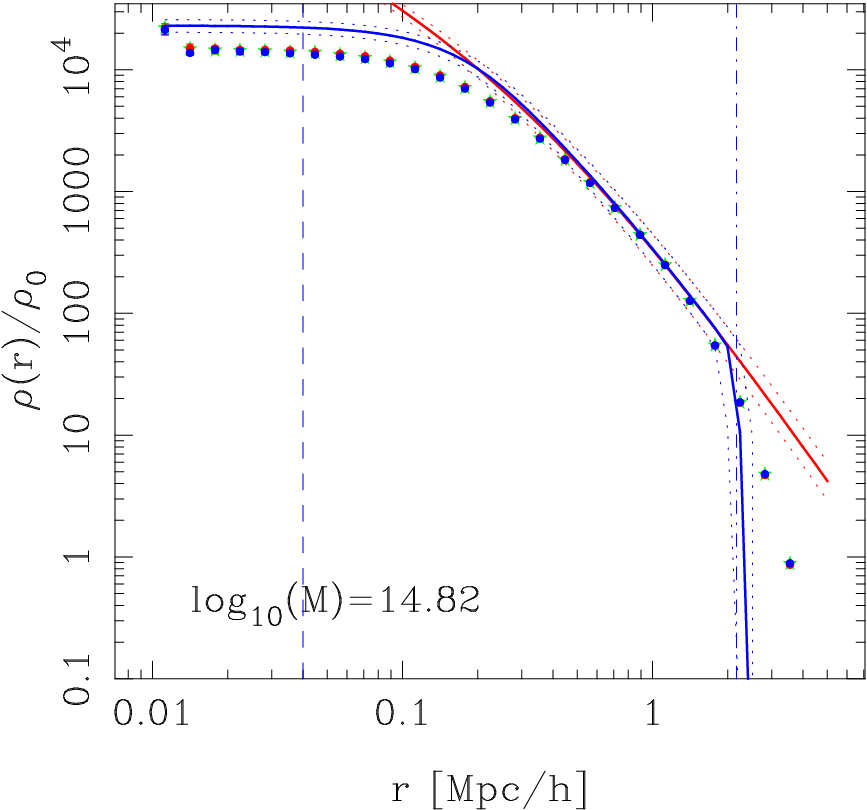}
  \includegraphics[width=5.1cm,clip=]{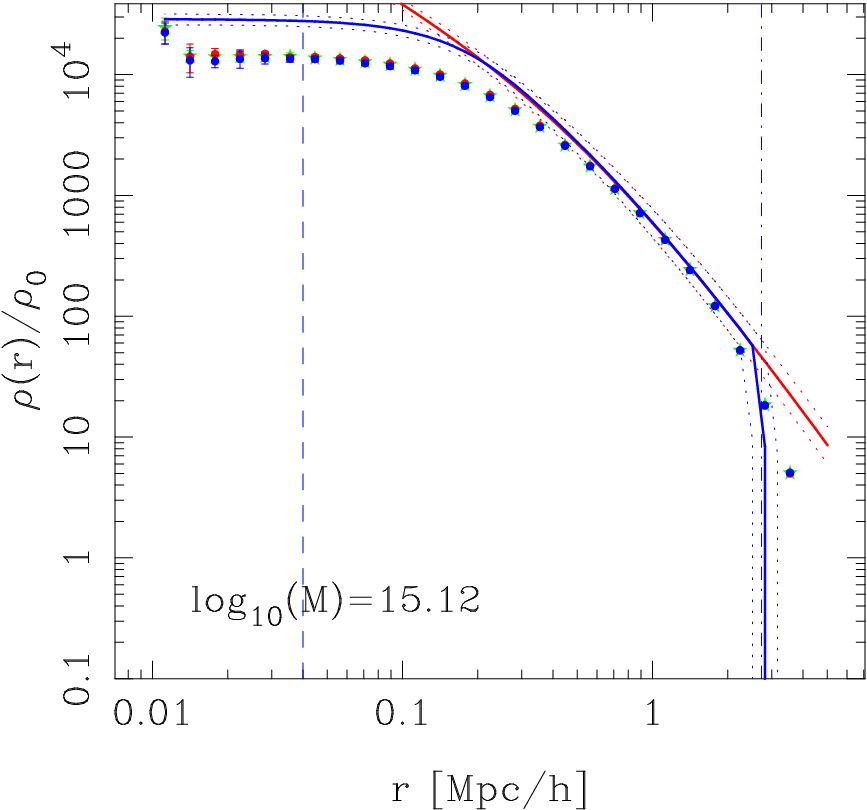}}
\caption{\small{Ensemble average density profiles of the dark matter
    haloes in the Gaussian ($\FNL=0$) simulations as a function of
    radius. The panels show haloes with bin centered masses
    $\log_{10}M/\Msol]\in\left\{13.62,14.22,14.92,15.42\right\}$.  In
    all panels: points with errors show estimates from the
    simulations; red solid lines give the results for the NFW model;
    blue solid lines give results for NFW model convolved with a
    Gaussian filter of radius 2.5 times the softening scale. The
    dotted lines either side of the main prediction, give the
    predictions if the halo mass associated with the upper and lower
    edges of the mass bin are used. Vertical dashed lines denote the
    softening length and vertical dot-dashed curves denote $r_{\rm
      vir}$ for the central mass of the bin. Note, the density
    profiles are constructed from only those partcicles that are
    members of the FoF groups.}\label{fig:Profile}}
\vspace{0.5cm}
\centering{
  \includegraphics[width=5.1cm,clip=]{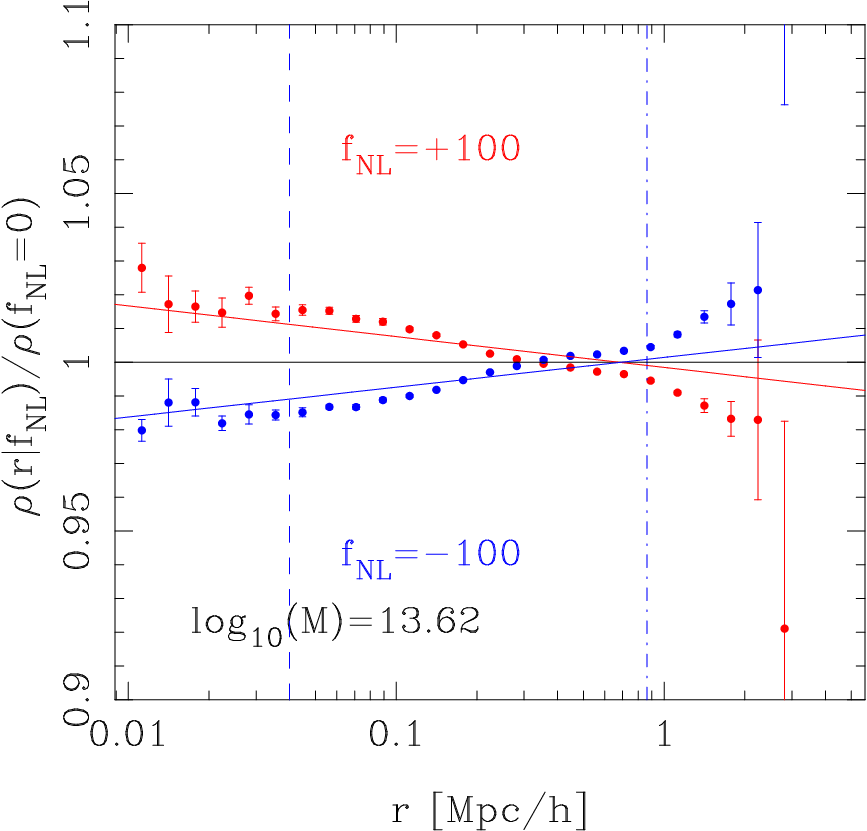}
  \includegraphics[width=5.1cm,clip=]{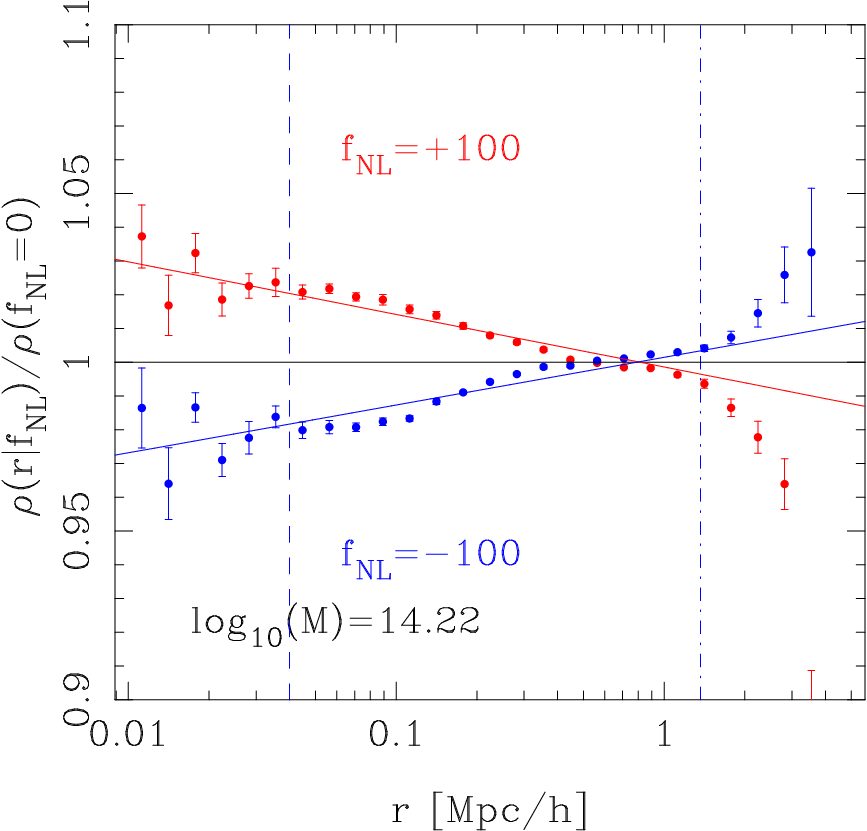}}\\
\centering{
  \includegraphics[width=5.1cm,clip=]{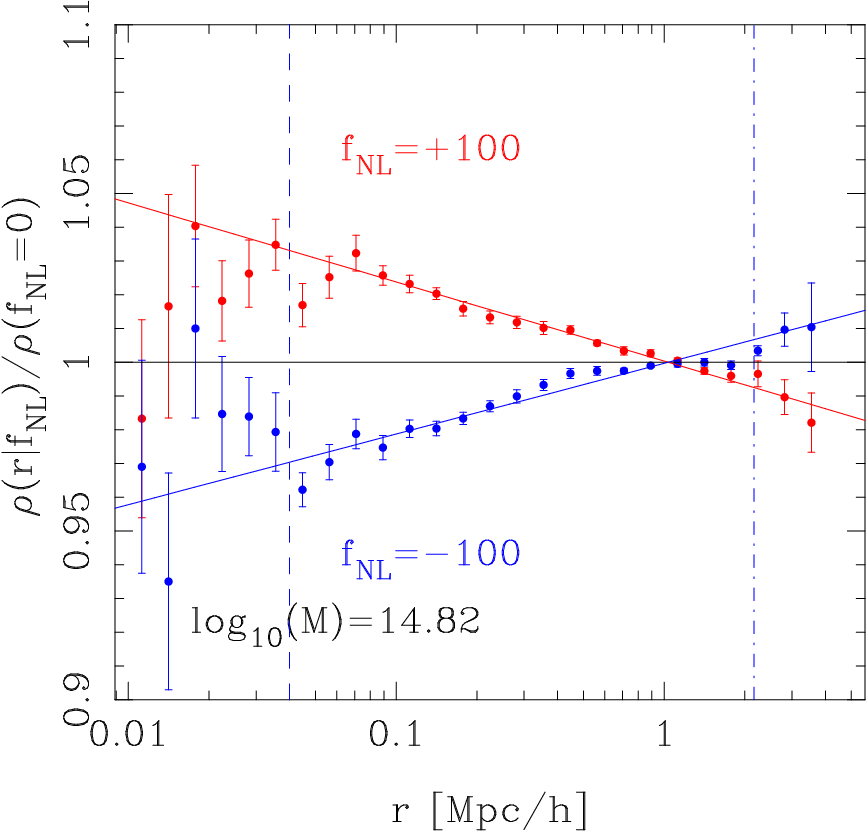}
  \includegraphics[width=5.1cm,clip=]{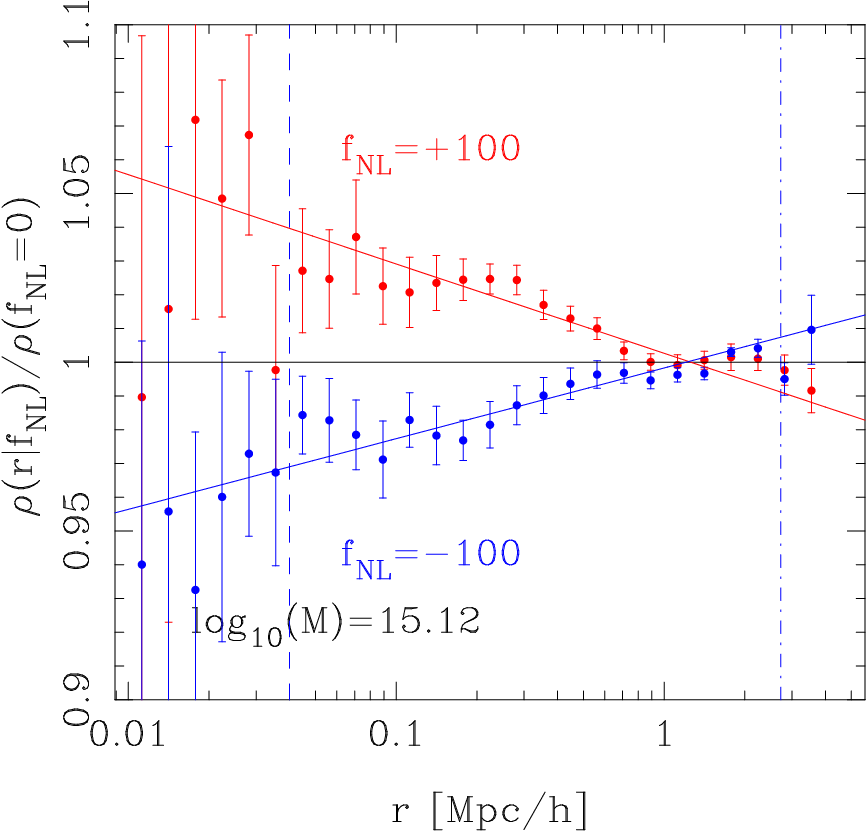}}
\caption{\small{Ensemble average of the ratio of the density profiles
    in the non-Gaussian model with the profiles in the non-Gaussian
    models as a function of radius. The red and blue colors
    corresponding to $\FNL=100$ and $\FNL=-100$, respectively. The
    points with errors denote estimates from the simulations; solid
    lines denote the log-linear profile ratio model as given by
    \Eqn{eq:RatioRho}, and the vertical lines are as in
    \Fig{fig:Profile}.}\label{fig:ProfileRatio}}
\end{figure*}


\subsection{Halo density profiles}

\subsubsection{Gaussian profiles}

The density profiles of dark matter haloes in simulations evolving
from Gaussian initial conditions has been studied in great detail.  A
reasonably good approximation for the spherically averaged density
profile is the Navarro, Frenk \& White model \cite{Navarroetal1997}.
This can be written:
\be \rho_{\rm NFW}(r|M)=\rhob\,\delta_{\rm c}(M) \left[
  \frac{r}{r_0}\left(1+\frac{r}{r_0}\right)^2 \right]^{-1} \ ,\ee
where the two parameters are the scale radius $r_0$ and the
characteristic density $\delta_c$. Note that if we define the halo
mass to be $M_{\rm vir}=4\pi r_{\rm vir}^3200\rhob/3$, then owing to
mass conservation there is only one free parameter: the concentration
parameter $c(M)\equiv r_{\rm vir}/r_0$, and we have
\be \delta_{\rm c}(M)=\frac{200c^3 /3}{\log\left(1+c\right)-c/(1+c)}\ .\ee
The parameter $c(M)$ can be obtained from the original model of NFW,
but instead we prefer to use the model of
\citet{Bullocketal2001}. Note that we correct $c(M)$ for the fact that
the definitions of the virial radius in the Bullock et al. model for
the concentration parameter and the Sheth \& Tormen mass definition
used for the mass function are different \citep[for details as to how
  to do this see][]{SmithWatts2005}.  Over the past decade a number of
alternative models for halo profiles have emerged. Owing to the
relatively low resolution of our haloes we believe that the original
model of NFW will be of sufficient accuracy to describe our haloes.

Figure \ref{fig:Profile} shows the ensemble average density profiles of dark
matter haloes in the simulations. The haloes were separated into a set
of mass bins of equal logarithmic width $\Delta\log_{10}M=0.3$, and
with the minimum halo mass from which a profile can be estimated being
taken as 50 particles ($M\sim1.5\times10^{13}\Msol$). Note that,
whilst the number of particles is relatively small for the lowest mass
haloes used in the profile estimation, as Table~\ref{tab:numbers}
shows, we are averaging over large numbers of haloes and multiple
simulations per mass bin. The figure shows the results for a
sub-sample of four of the mass bins, and these correspond to bins
$\{2,4,6,7\}= \left\{13.62,14.22,14.92,15.12\right\}
[\log_{10}M/\Msol]$ in Table~\ref{tab:numbers}.


\begin{table}
\caption{\label{tab:numbers}Expected number of haloes in the mass bins
  from which density profile averages are calculated, per
  simulation. Columns are: (1) number of mass bin; (2)--(3) lower and
  upper edges of the mass bin; (4)--(6) number of haloes in bin.}
\begin{ruledtabular}
\begin{tabular}{llllll}
Bin &  $\log_{10} M_1$\ & $\log_{10} M_2$\ & \# haloes & \# haloes   & \# haloes \\
 \# &  $[\Msol]$    &  $[\Msol]$   & $\FNL=0$ \ \    & $\FNL=100$ & $\FNL=-100$\\
\hline
1    &    13.169    &      13.469  &      617526.6  &      613869.9     &   620691.9 \\
2    &    13.469    &      13.769  &      311273.1  &      309856.2     &   312781.2 \\
3    &    13.769    &      14.069  &      144713.0  &      144526.5     &   144994.3 \\
4    &    14.069    &      14.369  &       60805.3  &       61050.2     &    60481.0 \\
5    &    14.369    &      14.669  &       21744.9  &       22108.4     &    21371.4 \\
6    &    14.669    &      14.969  &        5843.5  &        6080.4     &     5608.2 \\
7    &    14.969    &      15.269  &        1008.0  &        1092.5     &      922.5 \\
8    &    15.269    &      15.369  &          86.7  &         101.1     &       72.7 
\end{tabular}
\end{ruledtabular}
\end{table}


We estimate the density profile for each individual halo by taking
only the particles that are in the FoF halo ($b=0.2$). We compute the
halo center of mass and the radial distance of each particle from
this center. The particles are then binned into equal logarithmic
radial bins of thickness, $\Delta\log_{10} r[\Mpc]=0.1$. Our estimate
for the profile is then given by $\hat{\rho}(r_i,\bar{M})=
m_pN_{ij}[r_i,\bar{M}_j] /V_{\rm shell}(r)$, where
$N_{ij}[r_i,\bar{M}_j]$ is the number of particles in the $i$th radial
bin for haloes in mass bin $\bar{M}_j$ and the shell volume is $V_{\rm
  shell}=4\pi \left[(r_i+\Delta r_i/2)^3-(r_i-\Delta
  r_i/2)^3\right]/3$.

In \Fig{fig:Profile} we see that the model and the data do not agree
at the inner and outer parts of the profile. The data are
significantly flatter in the inner radius than the NFW model would
suggest. However, the softening length for the simulations was
$r\sim0.04\Mpc$, and so at this scale we expect the core to be
effectively of constant density. The vertical dash line in each plot
shows the softening length. The virial radius taken as $r_{\rm vir}$
is plotted for the mean mass in the bin and this is the vertical
dot-dash line in each panel. There is reasonable agreement between the
scale at which the FoF haloes are truncated and $r_{\rm vir}$.

We considerably improve the agreement between the NFW model and the
data by convolving the theoretical profiles with a Gaussian filter
function of radius 2.5 times the softening length, i.e.
\be {\tilde\rho}_{\rm NFW}(r|M)=
\int \frac{\dk}{(2\pi)^3}M U_{\rm NFW}(k|M) W(k) j_0(kr) \ , 
\label{eq:FourierProfile}\ee
where $W(k)\equiv \exp \left[-(2.5 l_{\rm soft}k)^2/2\right]$ and
where $U(r|M)\equiv \rho(r|M)/M$ is the mass normalized profile. We
truncate the profile at the virial radius and for the NFW model the
Fourier transform can be written as \citep{CooraySheth2002}:
\ba
f(c) U_{\rm NFW}(k|M) & = & -\frac{\sin(kcr_0)}{kr_0(1+c)}\nn \\
& & \hspace{-2.4cm}
\frac{}{}+\cos\left[kr_0(1+c)\right]\left\{C_i\left[kr_0(1+c)\right]
-C_i[kr_0]\right\}\nn \\
& & \hspace{-2.4cm} 
\frac{}{}+\sin\left[kr_0(1+c)\right]\left\{S_i\left[kr_0(1+c)\right]
-S_i[kr_0]\right\}, 
\label{eq:FourierProfileNFW}\ea
where $f(c)\equiv{[\log\left(1+c\right)-c/(1+c)]}$ and where $S_i$ and
$C_i$ are the standard sine and cosine integrals. This makes the
agreement between the low-mass halo samples and the data almost
perfect, however the higher mass halo samples would require larger
smoothing radii to explain the difference. One might motivate this by
the fact that for the highest mass bins these haloes are just forming
and as such are more likely to have complex structure, and so the
center of mass may not be a good proxy for the halo center. A better
choice may be the point of deepest potential. We shall not pursue this
matter further, but note that for the highest mass haloes in our
simulations the profiles appear to be flatter inside $r\sim0.5\Mpc$.


\begin{figure}
\centering{
  \includegraphics[width=6.5cm,clip=]{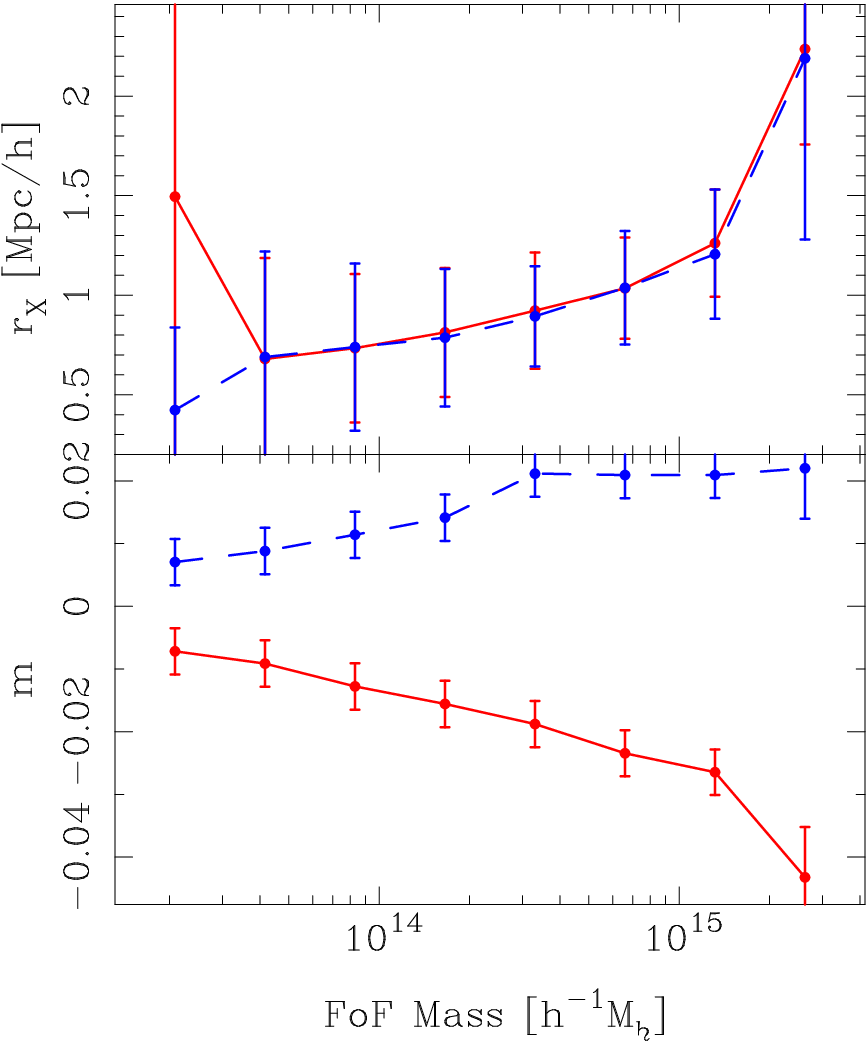}}
\caption{\small{Best fit parameters for the density profile ratio
    $R_{\rho}(\FNL,M)$, as defined in \Eqn{eq:RatioRho}, as a function
    of the FoF halo mass, measured at $z=0$. {\it Top panel:} the
    zero-crossing parameter $r_{\rm X}$. {\it Bottom panel:} the slope
    parameter $m$. The solid red lines denote the results for
    $\FNL=100$ and the blue dashed lines the results for
    $\FNL=-100$.}\label{fig:param}}
\end{figure}


\begin{figure}
\centering{
  \includegraphics[width=8cm,clip=]{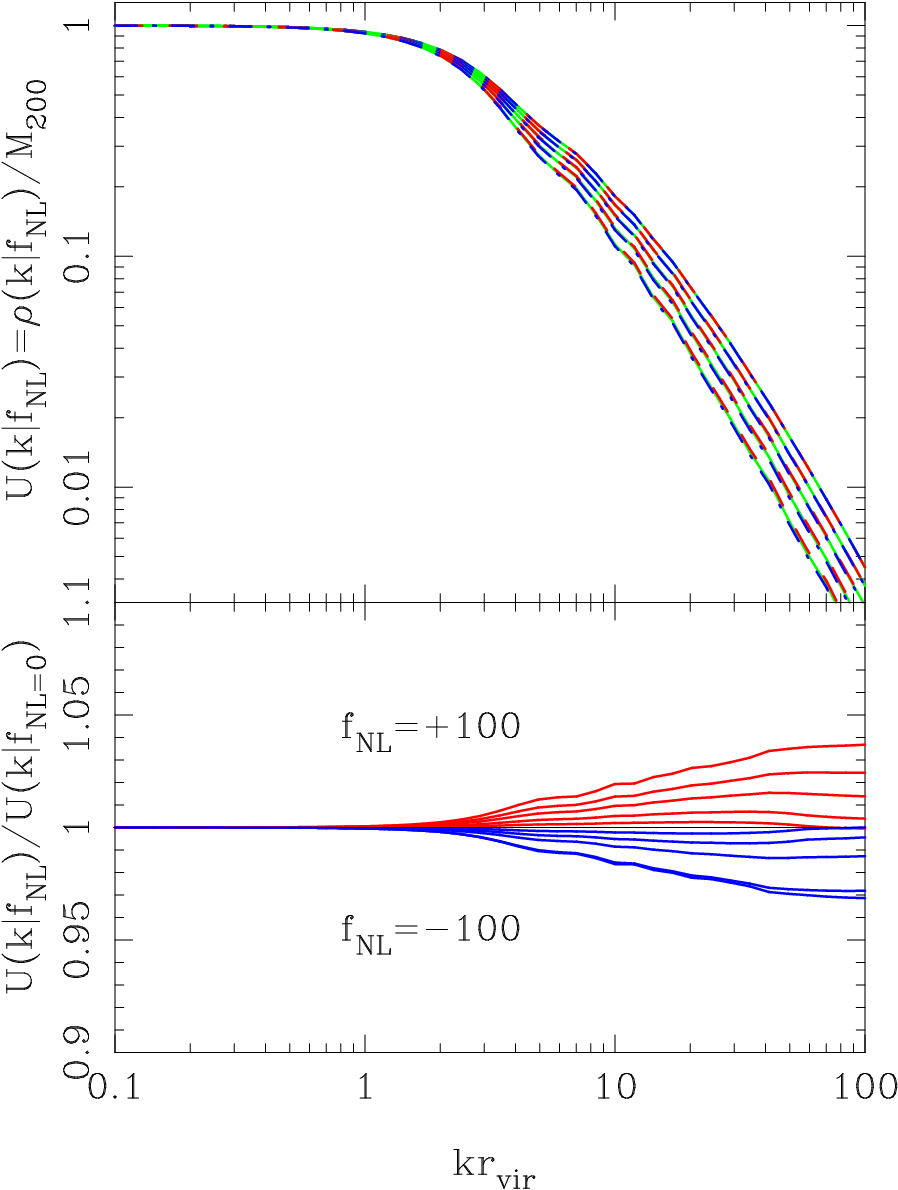}}
\caption{\small{Dependence of the Fourier transform of the mass
    normalized density profile on scale for several halo masses. {\it
      Top panel}: absolute value of the Fourier transform of the
    profile ($U(k|M)=\rho(k|M)/M$) as a function of wavenumber in
    units of inverse virial radius. The solid green, dashed red and
    dot-dashed blue lines show results for $\FNL\in[0,+100,-100]$. The
    curves from top to bottom show results for haloes with masses
    $M\in [5.0\times 10^{13}, 1.0\times 10^{14},5.0\times
      10^{14},1.0\times 10^{15},5.0\times10^{15}]\Msol$. {\it Bottom
      panel}: ratio of the non-Gaussian to Gaussian profiles in
    Fourier space. The higher mass haloes show stronger
    amplification/suppression for
    $\FNL=100/-100$.}\label{fig:FourierProfile}}
\end{figure}


\subsubsection{Primordial non-Gaussianity and density profiles}

The impact of PNG on the density profiles of dark matter haloes has
not been investigated previously. We therefore make a short
exploration.

\Fig{fig:ProfileRatio} shows the ratio of the density profiles in the
non-Gaussian simulations with those from the Gaussian ones. We
estimate this quantity by dividing the mean halo profile for each mass
bin for each $\FNL$ simulation with the corresponding one from the
Gaussian runs. The plotted points with errors are then the average and
1-$\sigma$ errors from the 12 simulations.  In the figure we clearly
see that there is an effect of PNG on the profiles. For the case of
$\FNL>0$, we find that the profiles are denser in the inner regions of
the halo; the converse is true for the case $\FNL<0$. For
$\FNL=\pm100$, the strength of the effect depends on the halo mass
considered: for cluster, group and small group mass haloes we find
effects of the order $\{\lesssim\pm4.5\%,\pm\lesssim\pm 3.5\%,\lesssim
\pm2.5\%\}$.

Since the power spectrum in the halo model depends on both the profile
and the square of the profile, these effects are important to
characterize for accurate clustering predictions on small scales. We
therefore attempt to model this in a simple way.  From
\Fig{fig:ProfileRatio} it can be seen that the profile ratio as a
function of log-radius appears to be almost a straight line. We
therefore fit a log-linear model to the measured profile
ratios. Explicitly the ratio model has the form:
\be 
R_{\rho}(r,M,\FNL)-1 = m \log_{10}
\left[r/r_{\rm X}\right] 
\label{eq:RatioRho}
\ee
where $m$ is the change in the slope and $r_{\rm X}$ is the zero
crossing scale, both of which depend on halo mass and $\FNL$. We fit
for the parameters $\{m,r_{\rm X}\}$ over the range of radii $(l_{\rm
  soft}<r<r_{\rm vir1})$, and the resultant model fits are shown in
\Fig{fig:ProfileRatio} as the solid lines.  The best fit parameters as
a function of halo mass are shown in \Fig{fig:param}, where we see
that, for all but the lowest mass bin in the $\FNL=100$ model, the
values of $r_{\rm X}$ increase with increasing mass and that these are
almost identical for both the positive and negative $\FNL$ models. On
the other hand, the values for the slope $m$ monotonically
decrease/increase for $\FNL$ positive/negative. Modulo the sign, these
values are similar for both $\FNL$ models.

To use this correction ratio in the halo model, we spline fit
$\{r_{\rm X},m\}$ as a function of mass, with the exception of the
lowest mass bin. We assume that the asymptotic limit for low masses is
$R_{\rho}\rightarrow 1$ as $M\rightarrow0$, and enforce this by adding
$\{r_{\rm X},m\}=\{0,0\}$ for $M=10^6\Msol$ as an extra data point in
the spline fitting.


What is actually required for the calculation of the power spectrum is
the mass normalized Fourier transforms of the density profiles, and we
show this in \Fig{fig:FourierProfile}. We calculate the Fourier
transform of the profile as in \Eqn{eq:FourierProfile}. Unfortunately,
owing to the radial dependent correction factor \Eqn{eq:RatioRho},
there is no analytic solution to this integral and so we compute this
numerically. We do however make the following improvement to
computational speed: the profile can be rewritten
\be U(k|M,\FNL)= U(k|M) R_{\tilde{\rho}}[k,M,\FNL]\ee
where $U(k|M)$ is given by \Eqn{eq:FourierProfileNFW}. We then
generate a bicubic spline \citep[see][for details]{Pressetal1992} fit
to the ratio $R_{\tilde{\rho}}$ as a function of $k r_{\rm vir}$ and
halo mass $M$. As a final note, the above expression is not normalized
correctly in that $U(k|M,\FNL)$ does not approach unity on large
scales. We are however free to renormalize $U(k|M,\FNL)$ and this can
be done through the operation:
\mbox{$\tilde{U}(k|M,\FNL)=U(k|M,\FNL)/U(k=0|M,\FNL)$}.


\begin{figure*}
\centering{
  \includegraphics[width=8.5cm,clip=]{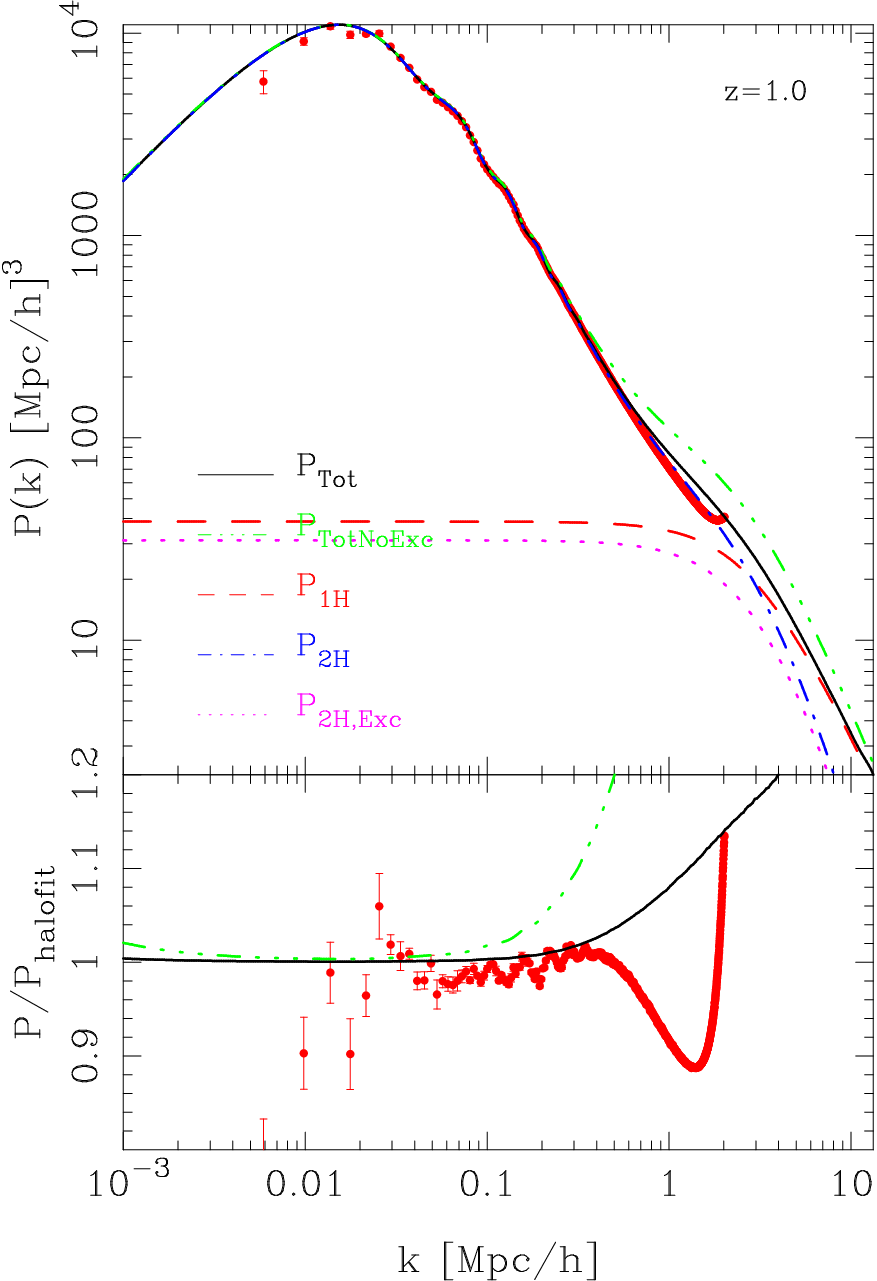}\hspace{0.2cm}
  \includegraphics[width=8.5cm,clip=]{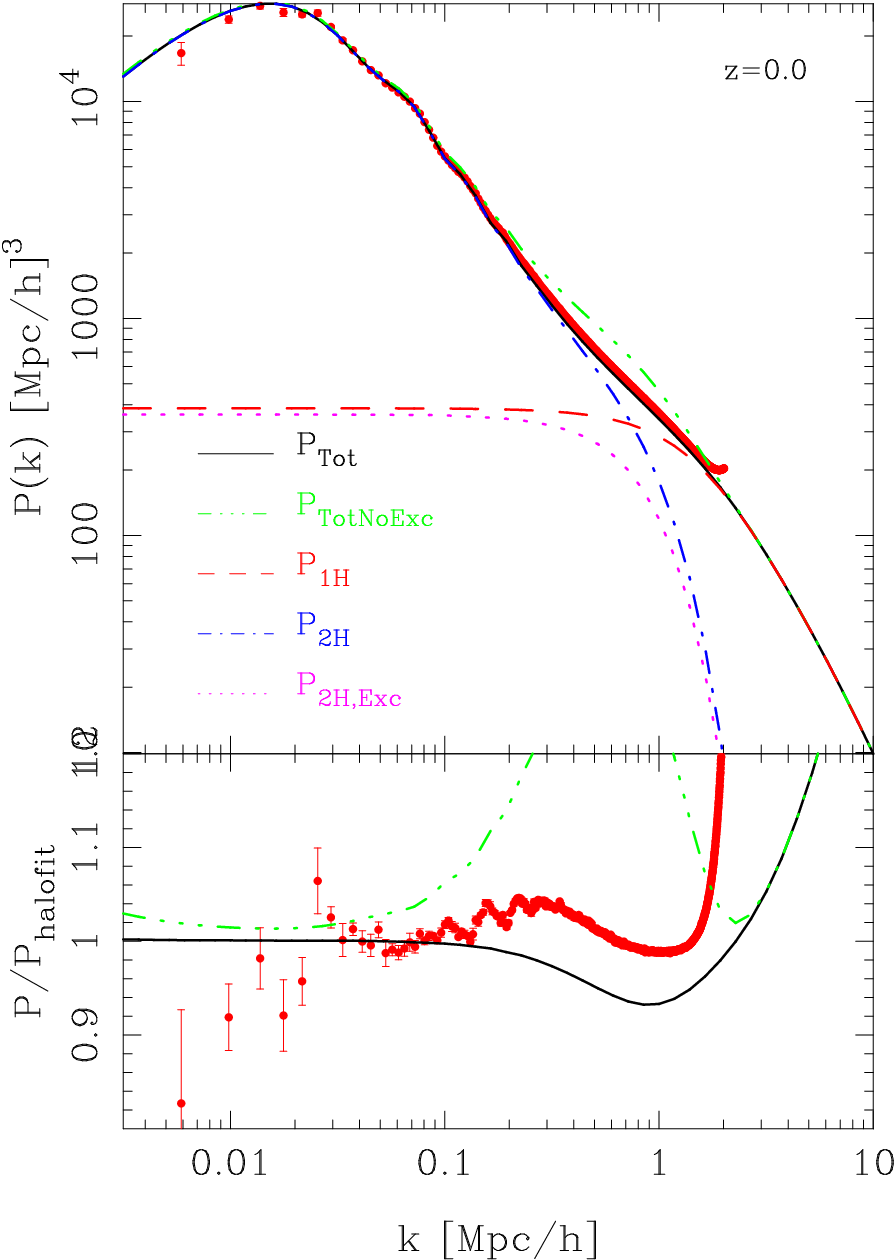}}
\caption{\small{Comparison of the matter power spectrum in the halo
    model and measurements from an ensemble of numerical simulations
    (see \S\ref{sec:simulations} for details). {\it Top panel}:
    Absolute power. Points with error bars show estimates from the
    simulations. The halo model predictions are: red dash line denotes
    $P_{\1H}$; blue dot-dash line denotes $P_{\2H}$; magenta dotted
    line represents $P_{\2H}^{\rm Exc}$; the black solid line shows
    the total halo model prediction including subtraction of the halo
    exclusion term; the green triple-dot-dash curve is the same, but
    neglecting the halo exclusion term. {\it Bottom panel}: ratio of
    the matter power spectra with predictions from {\tt
      halofit}\citep{Smithetal2003}. Points and line styles are as
    above.  }\label{fig:GaussPow} }
\end{figure*}


\begin{figure*}
\centering{
  \includegraphics[width=8cm,clip=]{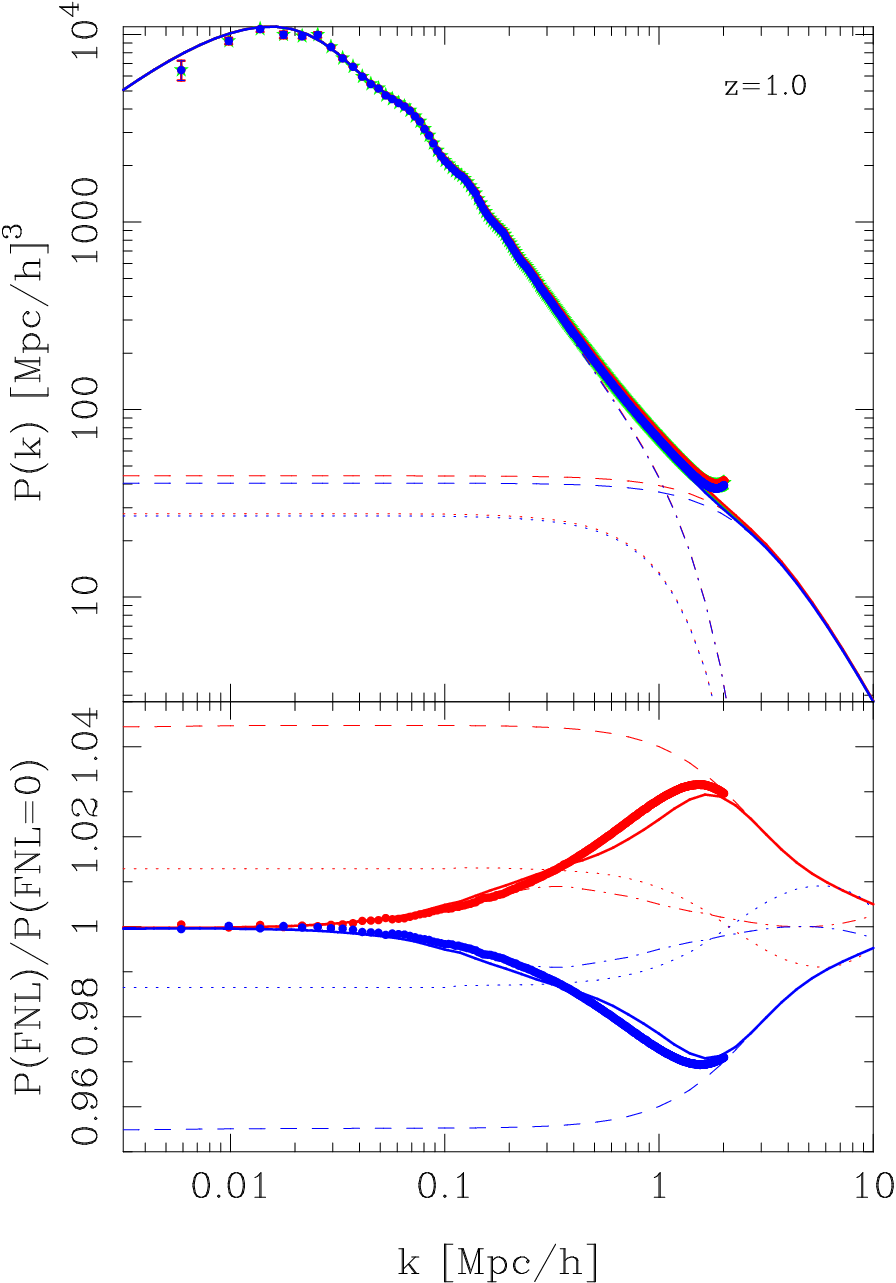}\hspace{0.3cm}
  \includegraphics[width=8cm,clip=]{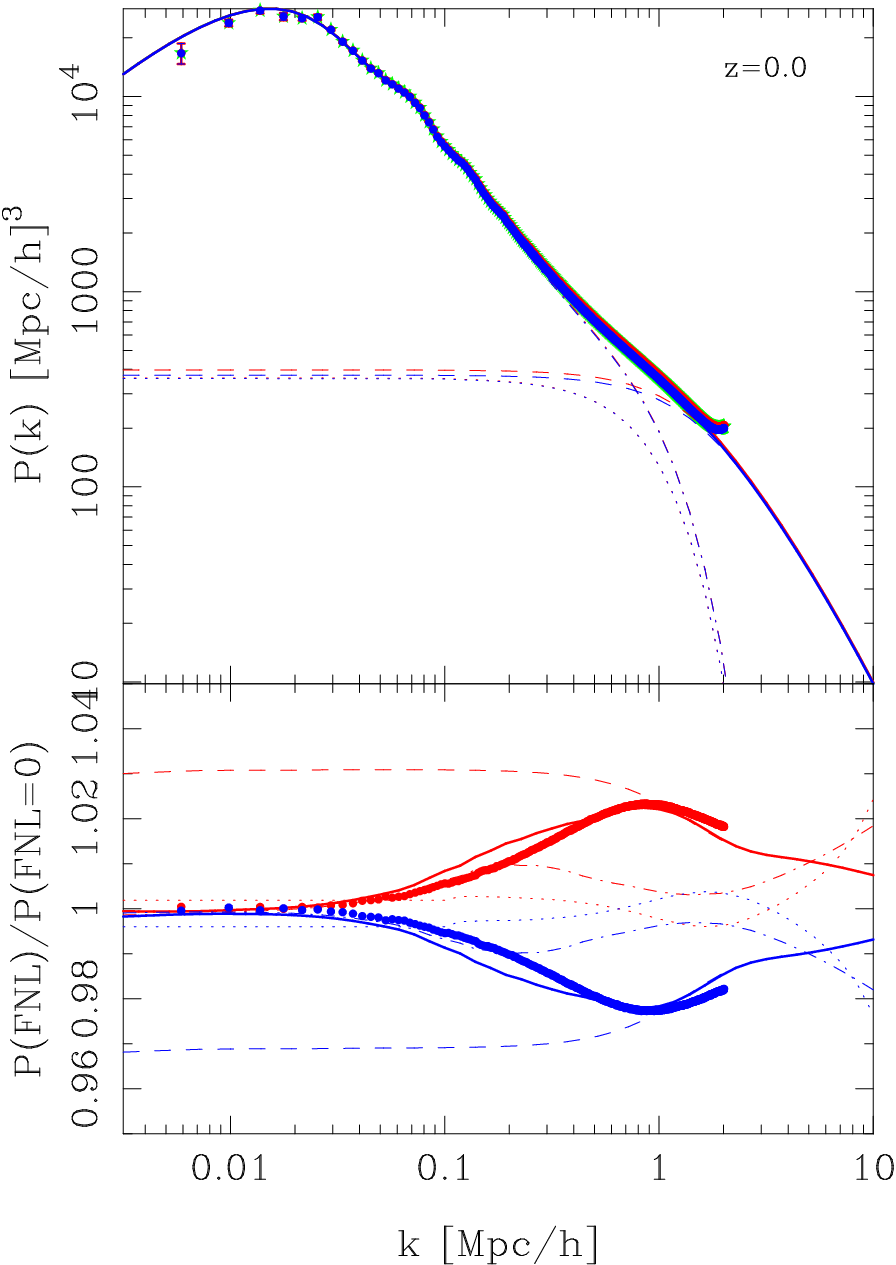}}
\caption{\small{Comparison of the matter power spectrum in models with
    Gaussian and non-Gaussian initial density fluctuations at
    redshifts $z=1.0$ (left sub-fig.) and $z=0.0$ (right sub-fig.).
    {\it Top panels:} Absolute power. Points with error bars show
    results for the simulations and the colors green, red and blue
    denote the models $\FNL=\{0,+100,-100\}$. The lines represent halo
    model predictions: dash lines denote $P_{\1H}$; dot-dash lines
    denote $P_{\2H}$; dotted lines denote $P_{\2H}^{\rm Exc}$; the
    solid line represents the total halo model prediction including
    subtraction of the halo exclusion term. {\it Bottom panels}: ratio
    of the matter power spectra in the $\FNL=+100$ and $-100$ models
    with respect to the Gaussian ($\FNL=0$) results. Points and line
    styles are as above.}\label{fig:PNGPow}}
\end{figure*}


\begin{figure*}
\centering{
  \includegraphics[width=8cm,clip=]{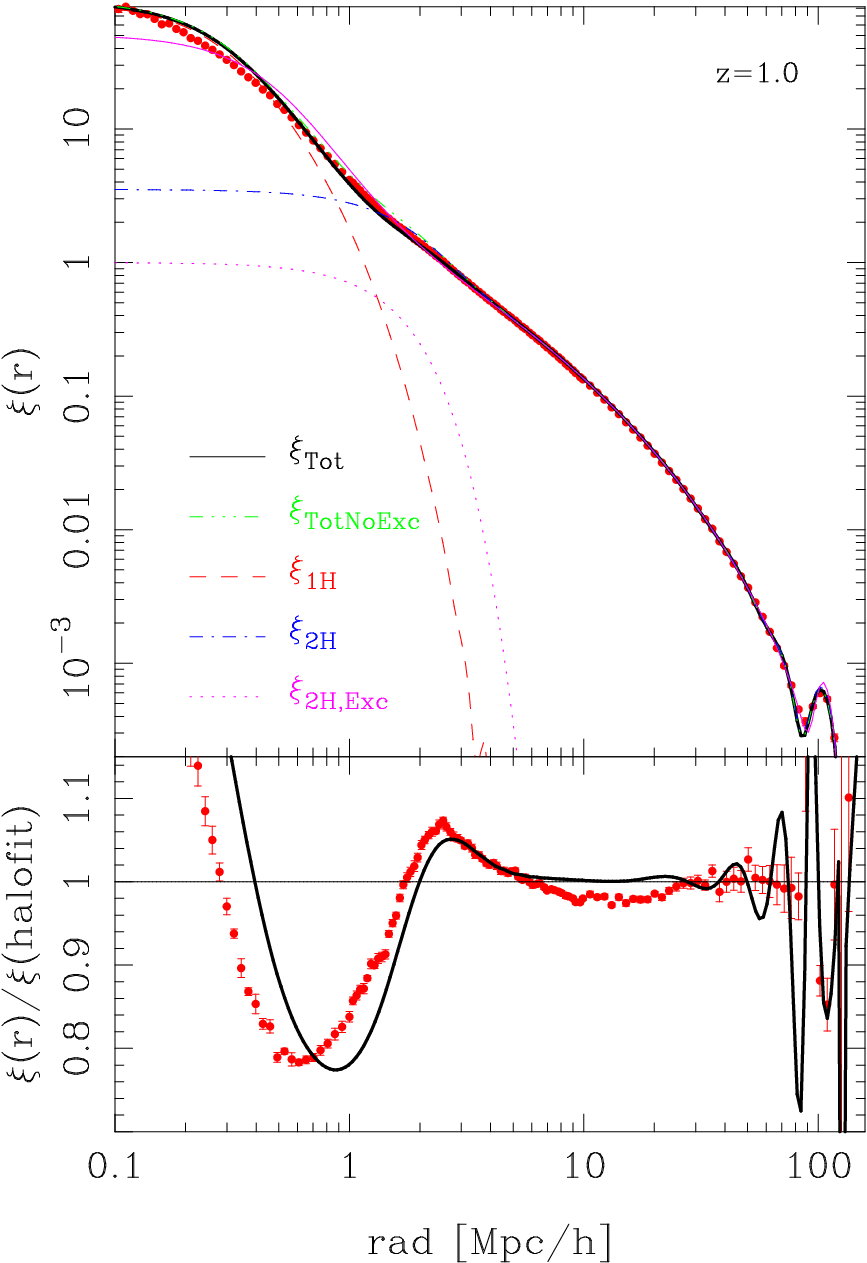}\hspace{0.3cm}
  \includegraphics[width=8cm,clip=]{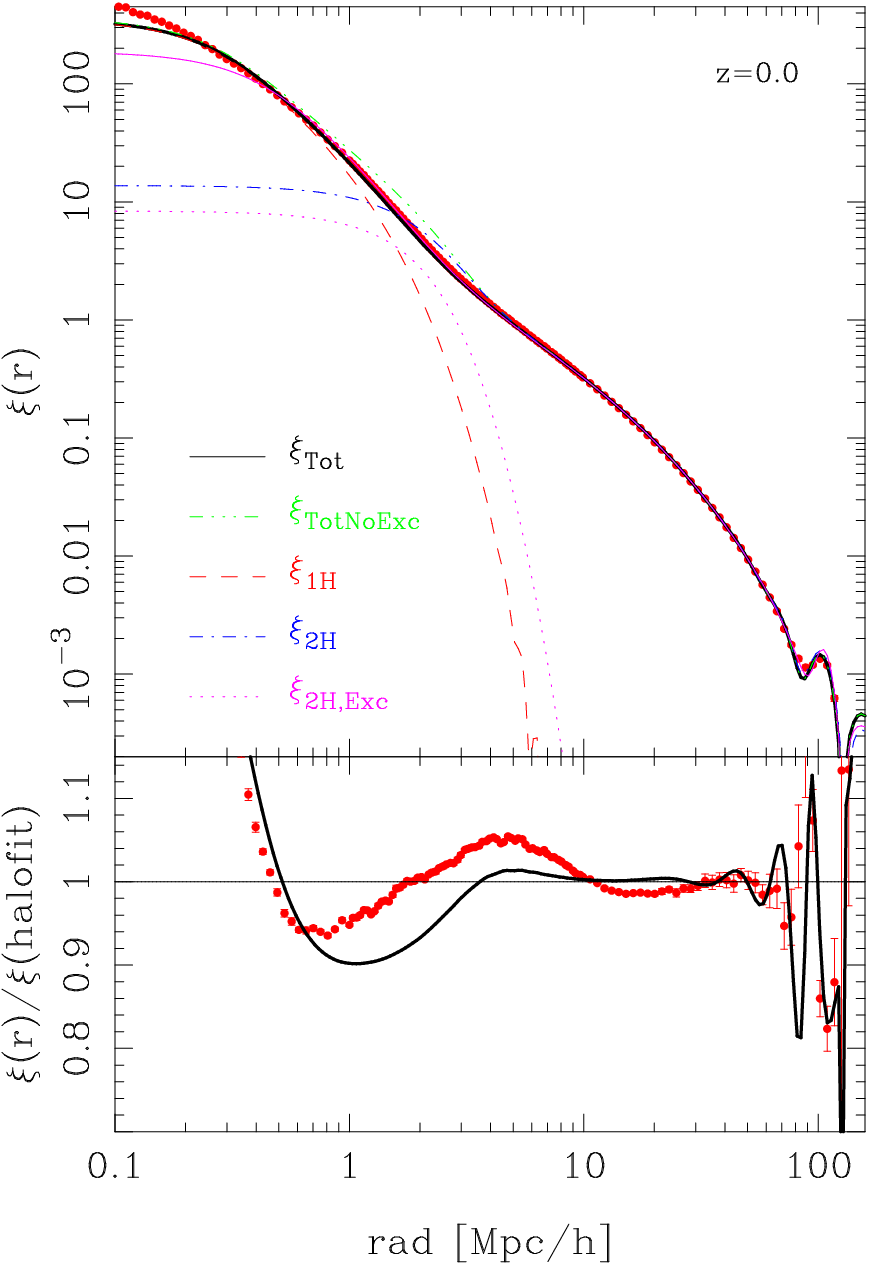}}
\caption{\small{Matter correlation function as a function of spatial
    scale measured in the Gaussian simulations at redshifts $z=1.0$
    (left sub-fig.) and $z=0.0$ (right sub-fig.). {\it Top panels}:
    absolute correlation function. Points with errors denote the
    measurements from the simulations and the lines denote the Halo
    Model predictions: dash lines denote $\xi_{\1H}$; dot-dash lines
    denote $\xi_{\2H}$; dotted lines denote $\xi_{\2H}^{\rm Exc}$;
    solid lines represent the total halo model prediction including
    subtraction of the halo exclusion term; triple dot-dash lines show
    the same but without subtracting the exclusion term. {\it Bottom
      panel}: ratio of the measurements and halo model predictions
    with respect to the nonlinear correlation function from {\tt
      halofit}.}\label{fig:CorrGauss}}
\end{figure*}


\begin{figure*}
\centering{
  \includegraphics[width=8cm,clip=]{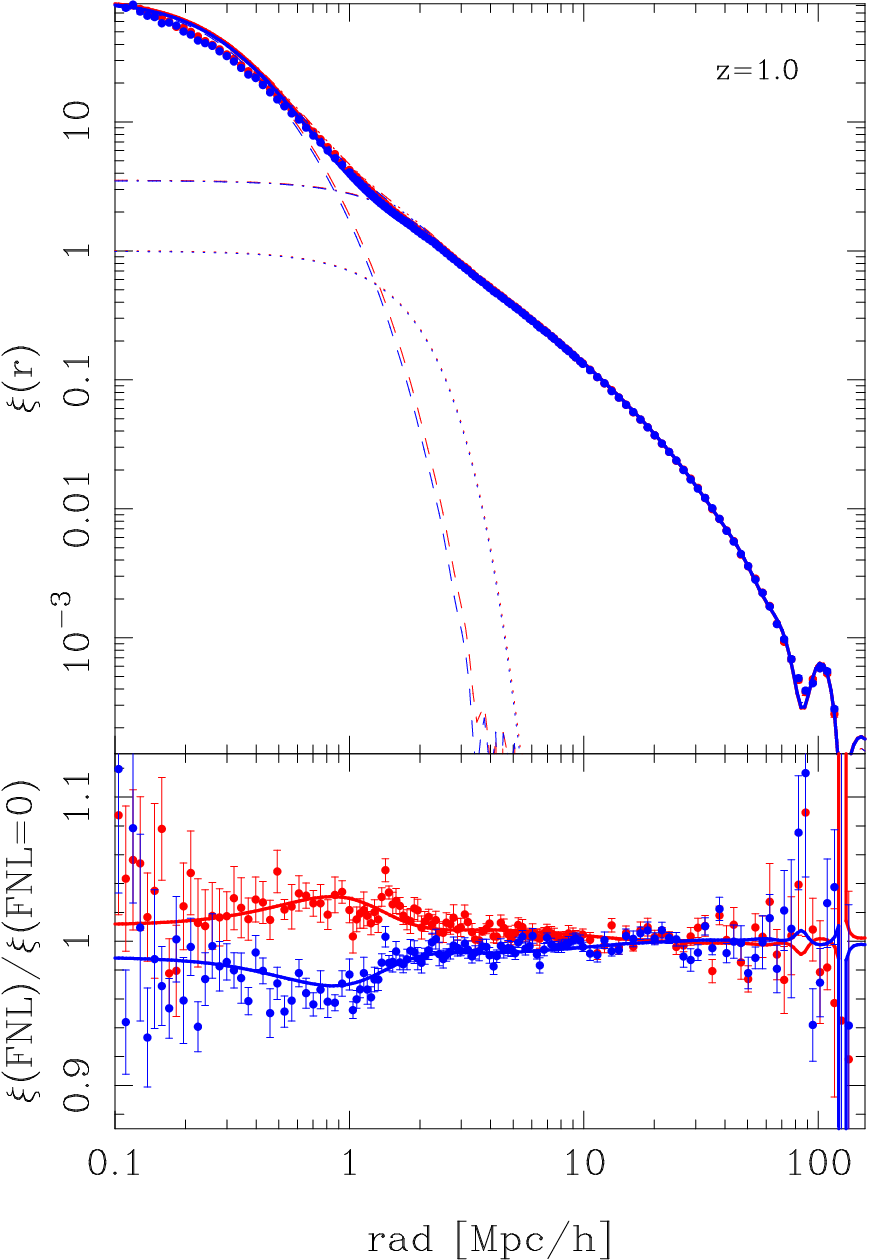}\hspace{0.3cm}
  \includegraphics[width=8cm,clip=]{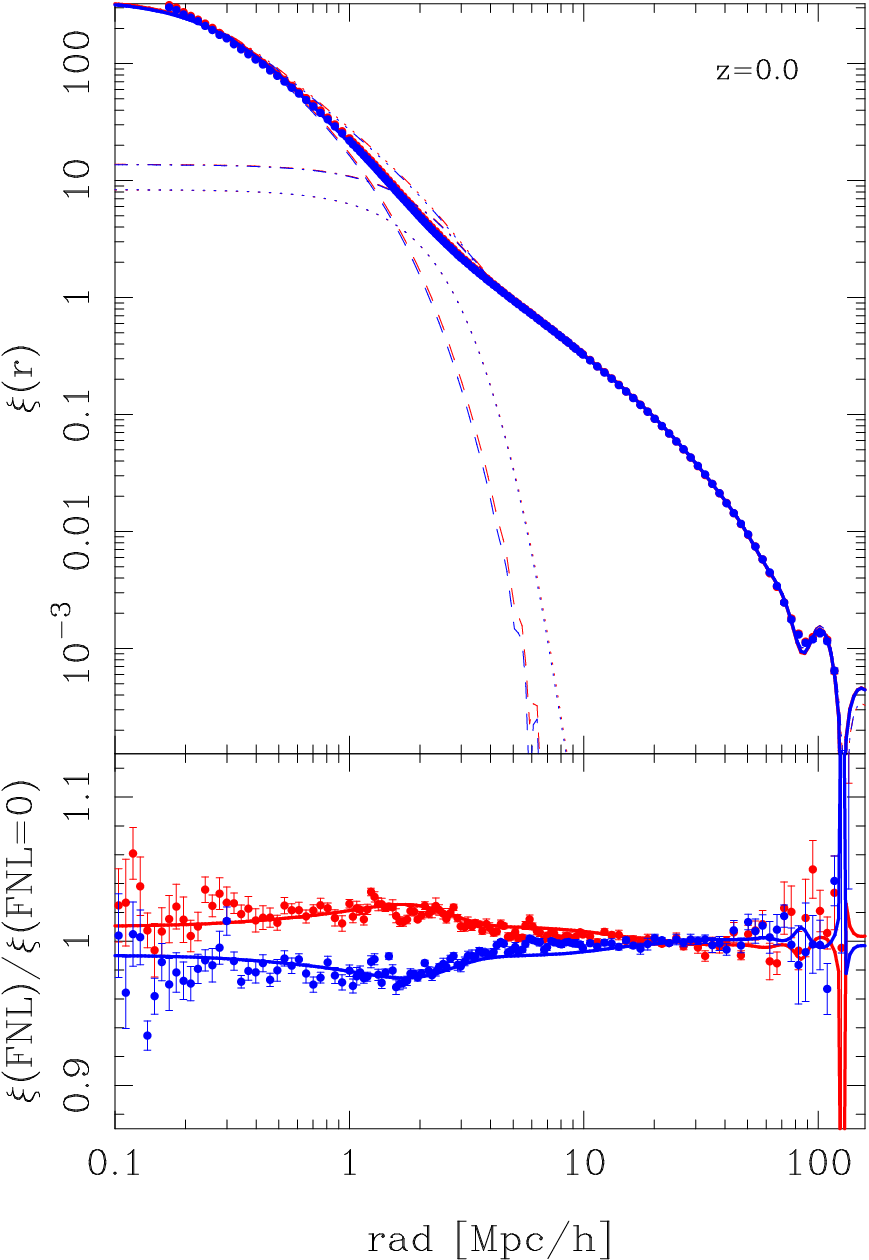}}
\caption{\small{Comparison of the matter correlation function in
    models with Gaussian and non-Gaussian initial density fluctuations
    at redshifts $z=1.0$ (left sub-fig.) and $z=0.0$ (right sub-fig.)
    {\it Top panels}: absolute correlation function. Points with
    errors show results for the simulations and the colours green, red
    and blue denote the models $\FNL=\{0,+100,-100\}$. The lines
    represent halo model predictions: dash lines denote $\xi_{\1H}$;
    dot-dash lines denote $\xi_{\2H}$; dotted lines denote
    $\xi_{\2H}^{\rm Exc}$; the solid line represents the total halo
    model prediction including subtraction of the halo exclusion
    term. {\it Bottom panels}: ratio of the matter correlation
    functions with $\FNL=+100$ and $-100$ with respect to the Gaussian
    ($\FNL=0$) results. Points and line styles are as
    above.}\label{fig:CorrPNG}}
\end{figure*}


\section{Results}\label{sec:results}

We now put together all of the components of the halo model and make
predictions for the nonlinear matter power spectrum and the matter
correlation function in the Gaussian case and then for the models with
local PNG. A practical note on evaluating the 2-Halo term for the mass
distribution: we are required to compute integrals over an infinite
domain in halo mass. This is computationally challenging, and so
instead we make the following approximation:
\ba P_{\2H} & \rightarrow & P_{\rm NL}\left[\frac{1}{\tilde{\rhob}}
  \int_{M_1}^{M_2} dM n(M) M b(M)
  U(k|M)\right]^2,\ \label{eq:Pow2Hmod} \ea
where $\tilde{\rho}=\int_{M_1}^{M_2} dM n(M) M b(M)$. Provided $M_1$
and $M_2$ are sufficiently small and large halo masses, then the above
integral approaches the exact answer of infinite limits over a
restricted range of $k$. We set
$\{M_1,M_2\}=\{1.0\times10^6,1.0\times10^{16}\}\Msol$.


\subsection{Estimating the power spectrum}\label{ssec:estpower}

The density Fourier modes were estimated using the conventional Fast
Fourier Transform (FFT) method: the dark matter particles were
assigned to a regular cubical grid using the `Cloud-In-Cell' (CIC)
scheme \cite{HockneyEastwood1988}, and throughout we used $N_{\rm
  g}=1024^3$ sized Fourier meshes. The FFT of the gridded density
field was then computed using the {\tt FFTW} routines
\cite{FFTW}. Each resulting Fourier mode was corrected for the
convolution with the mesh by dividing out the Fourier transform of the
mass-assignment window function. For the CIC algorithm this
corresponds to the following operation:
\be 
\delta_{\rm d}(\bk)=\frac{\delta_{\rm g}(\bk)}{W(\bk)} \ ;
\hspace{0.3cm}
W(\bk)=\prod_{i=1}^3\left\{\left[\frac{\sin 
\left[\pi k_i/2k_{\rm Ny}\right]}{\left[\pi k_i/2k_{\rm Ny}\right]}\right]^2\right\}\ ,\ee
where sub-script d and g denote discrete and grid quantities, and
where $k_{\rm Ny}=\pi N_{\rm g}/L$ is the Nyquist frequency.

The discrete power spectra on scale $k_l$ are then estimated by
performing the following sums,
\be \widehat{\overline{P}}_{\rm
  d}(k_l)=\frac{\Vu}{N_k}\sum_{l=1}^{N_k}\left|\delta_{\rm
    d}(\bk_l)\right|^2 \, ,\label{eq:powestimate}\ee
where $N_{k}$ is the number of Fourier modes in a spherical shell in
$k$-space of thickness $\Delta k$. 


\subsection{Matter power spectrum: Gaussian case}

Figure~\ref{fig:GaussPow} presents the nonlinear matter power spectra
measured in the simulations for the Gaussian case at redshifts $z=1.0$
and $z=0.0$, left and right panels respectively. The top sections of
the panels show the absolute power and the lower ones the ratio with
respect to the predictions from {\tt halofit}\cite{Smithetal2003}. The
figure also shows a term by term breakdown of the halo model
predictions. It can clearly be seen that the sum $P_{\1H}+P_{\2H}$
without subtracting the exclusion term over shoots the measured
nonlinear power by a factor of 2-3 (green triple dot dash line). The
subtraction of the term due to halo exclusion
$P_{\1H}+P_{\2H}-P_{\2H}^{\rm Exc}$ significantly improves the
predictions of the model on all scales and at both of the redshifts
considered. We see that at $z=0.0$ the total halo model result is
within a few percent of the measured power from the numerical
simulations on all scales measured. At $z=1.0$ the same statement is
true except on small scales, $k>0.5\kMpc$ where there appears to be a
significant disagreement of the order 20\%, this is partly due to the
failure of {\tt halofit} on these scales (which is the input nonlinear
power spectrum for the 2-Halo term). Improving the modeling of the
halo-halo center clustering will most likely solve this issue. This
discrepancy may also be alleviated by the fact that the Nyquist
frequency for the power spectra is $k_{\rm Ny}=\pi 1024 /1600 \sim
2.0\kMpc$, and that one should only really trust the results up to
$k_{\rm Ny}/2\sim1.0\kMpc$.

On large scales we see that the $P_{\1H}$ term (dashed red line)
asymptotes to a constant value $\sim \{40, 400\} \Mpccube$ at $z=1$
and $z=0.0$, respectively. This is significantly larger than the
expected amplitude due to the shot-noise of the particles
$1/\bar{n}=V/N\sim3.8 \Mpccube$. The term $P_{\2H}^{\rm Exc}$ (magenta
dotted line) effectively kills the excess shot-noise of the 1-Halo
term. This is entirely consistent with the theoretical expectations in
Appendix~\ref{sec:RPTHM}. The subtraction of the term due to halo
exclusion therefore is an essential correction to make in order to
make realistic predictions in the halo model.


\subsection{Matter power spectrum: PNG case}

Figure~\ref{fig:PNGPow} shows the effects of PNG on the matter power
spectrum at redshifts $z=1$ and $z=0$, left and right sub-figures
respectively. The measurements with errors are the ensemble average
power obtained from the 12 realizations per $\FNL$.  The predictions
from the halo model calculation are also plotted and we see good
agreement.  However, the difference between the $\FNL=\{0,+100,-100\}$
models is very small and so we take the ratio of the PNG models with
respect to the Gaussian case, and this is what is plotted in the lower
sections of the figures. This clearly shows, as was seen earlier in
\Fig{fig:PT}, that the changes are of the order $\sim3.5\%$ at $z=1.0$
and of the order $\sim2.5\%$ at $z=0.0$.  The halo model predictions
for this ratio are in excellent agreement with the measured ratios.

The strength of the effect appears to peak around $k\sim1$, and then
declines at higher $k$. This can be understood in the following way:
as one goes to higher $k$ the 1-Halo term comes to dominate. The
integrand for this is $n(M)M^2U(k|M)$, and for cluster mass scales it
peaks around $k\sim1$ and then decays strongly, whereas for group and
galaxy mass scales it peaks at much higher $k$.  As we have seen
earlier in \S\ref{sec:ingred}, the effect of PNG on the mass function
and density profiles is strongest for the highest mass haloes. This
then leads us to expect that the effect of PNG on the nonlinear power
spectrum should peak around $k\sim1$. The dashed lines in
\Fig{fig:PNGPow} show the 1-Halo term and so give confirmation of this
logic.


\subsection{Matter correlation function: Gaussian case}

We estimate the matter correlation function for the $z=1$ and $z=0$
snapshots in each simulation to an accuracy of order 3-5\% using our
fast and exact correlation code {\tt DualTreeTwoPoint}, which is based
upon the $k$D-tree data structure, and the code is parallelized using
MPI calls. Thus on averaging over the 12 simulations we expect results
that are accurate to $5\%/\sqrt{12-1}\lesssim2\%$.

Figure~\ref{fig:CorrGauss} presents the ensemble average estimate of
the matter correlation function in the Gaussian models over three
decades in spatial scale at redshifts $z=1.0$ and $z=0.0$, left and
right panels respectively. The figure also shows the halo model
predictions appear in remarkably good agreement with the simulation
data. The exact deviations are hard to quantify on the log-scale and
so we take the ratio of the theory and simulation measurements with
respect to the {\tt halofit} model correlation function. 

We now see that the halo model predictions are better than 10\% over
the entire range of scales and redshifts considered. The predictions
are somewhat worse at the 2-- to 1--Halo cross-over scale
(i.e. $r\in[2,10]\Mpc$), also on the very largest of scales around the
BAO feature and on the smallest scales $r\lesssim0.2\Mpc$.  We
emphasize that none of the halo model parameters were tuned to fit the
clustering statistics directly.

In the figure we also show the result for the halo model calculation
if no exclusion correction is made, and we see that predictions
significantly overshoot the measurements by factors of a few on small
scales, especially at low redshift. The figure also shows that the
exclusion correction essentially kills the contribution from the
2-Halo term to the correlation function on small scales. Furthermore
this correction also kills some of the contribution of the 1-Halo term
to the correlation function on scales larger than $r\sim2\Mpc$.


\subsection{Matter correlation function: PNG case}

In Figure~\ref{fig:CorrPNG} we present the ensemble average estimate
of the matter correlation function in the models evolving from PNG
initial conditions at redshifts $z=1.0$ and $z=0.0$, left and right
panels respectively. As for the correlation function in the Gaussian
case the halo model predictions with exclusion provide a remarkably
good description of the clustering. The differences are not clearly
visible on the log-scale, and so we take the ratio of the PNG models
with the Gaussian case. Note that we construct this ratio for each
simulation and compute the ensemble average.

The lower panels of the figures show that, as in the case for the
matter power spectrum, there is a significant signal of $\FNL$ on the
small-scale correlation function. At $z=0.0$ the signal is of the
order $\sim2.5\%$ and affects the clustering on scales $r<5\Mpc$. At
higher redshift, $z=1$, we clearly see the same general trend but the
relative difference in the signal is a little larger $\sim3-4\%$.
However, the measurements appear slightly noisier. 

Once again the halo model predictions with exclusion provide an
excellent description of the ratio, being accurate to $\sim1\%$
precision.


\section{Conclusions}\label{sec:conclusions}

In this paper we have developed the halo model of large-scale
structure for application to computing matter clustering statistics in
models with PNG initial conditions. In particular, we considered the
case of local quadratic corrections to the primordial potential,
characterized by the parameter $\FNL$, see \S\ref{sec:ng} for details.

In \S\ref{sec:simulations} we provided details of the large ensemble
of $N$-body simulations that we employed in this study.  

In \S\ref{sec:modelling} we explored standard nonlinear perturbation
theory techniques to predict the matter power spectrum in models with
PNG.  It was demonstrated that the next-to-leading-order correction to
the power spectrum worked very well up to scales $k<0.2\kMpc$ for the
ratio with respect to the Gaussian case, but that on smaller scales
the PT failed to reproduce the simulation results. Further, the
absolute power was only well reproduced by the PT on scales
$<0.1\kMpc$.

In \S\ref{sec:modelling}, we reviewed the halo model and gave a
calculation of halo exclusion. In Appendix~\ref{sec:RPTHM} we also
showed theoretically that exclusion can help resolve the problem of
the large-scale excess power in the halo model.

In \S\ref{sec:ingred} we performed a numerical study of the
ingredients of the halo model, in the context of PNG.  We studied the
halo mass function; the halo bias and the density profiles. For the
mass function ratios, we confirmed that the existing models of
\citet{LoVerdeetal2008} and \citet{Matarreseetal2000} were in
excellent agreement with the simulations, once the peak height was
rescaled \citep{Grossietal2009}. We found that the Gaussian mass
function of \citet{ShethTormen1999} was a poor description of the
simulation data.

In \S\ref{sec:ingred} we examined the halo bias in the context of PNG
and showed that, if the halo model is correct, then there must be a
small asymetry in the scale-dependence of the bias in non-Gaussian
models on large scales.

In \S\ref{sec:ingred} we explored the density profiles of dark matter
haloes in the context of PNG. This has not been undertaken before and
we showed that halo profiles become more (less) dense in the presence
of $\FNL>0$ ($\FNL<0$). We found, for $\FNL=\pm100$, that cluster,
group and small group mass haloes were modified at the level of
$\{\lesssim\pm4.5\%,\pm\lesssim\pm 3.5\%,\lesssim \pm2.5\%\}$. We
modeled these effects with a simple log-linear model.

In \S\ref{sec:results} we presented the two-point clustering
statistics of the matter in the simulations and in the halo model.  We
showed that including halo exclusion in the model was important in
order to produce accurate results. We also demonstrated that on large
scales the halo exclusion term almost exactly canceled the excess
large-scale power arising in the 1-Halo term. This appears to solve a
long standing technical problem for the halo model.

For the case of the absolute power spectra, whilst the theory and the
measurements differed by up to 10\%, the ratio of the non-Gaussian to
Gaussian predictions and measurements differed by of the order
$\sim1\%$. For the case of the correlation functions, the absolute
predictions on small scales were very good and the ratio of the
non-Gaussian to Gaussian predictions and measurements were also
accurate to $\sim1\%$

We conclude that the modeling that we have developed in this paper,
will be good enough for predicting the absolute value of matter
clustering statistics to within $\lesssim10\%$ and their ratios to
$\lesssim1\%$. We anticipate that this will be useful for constraining
$\FNL$ from measurements of the nonlinear shear correlation function
in future weak lensing surveys such as EUCLID \citep{EUCLID2010} and
LSST \citep{LSST2009}.


\section*{Acknowledgements}
We thank Adam Amara, Tsz Yan Lam, Cristiano Porciani, Roman
Scoccimarro, Emiliano Sefusatti, Ravi Sheth and Uros Seljak for useful
discussions.  We kindly thank V. Springel for making public {\tt
  GADGET-2} and for providing his {\tt B-FoF} halo finder. RES and VD
would like to acknowledge the hospitality of the ``Centro de Ciencias
de Bensque -- Pedro Pascual'' where some of this work was performed.
RES acknowledges support from a Marie Curie Reintegration Grant and
the Alexander von Humboldt foundation. RES and VD were partly
supported by the Swiss National Foundation under contract
200021-116696/1 and WCU grant R32-2008-000-10130-0. VD was also partly
supported by the University of Zurich under contract FK UZH
57184001. LM was supported by the Deutscher Forschungsgeminschaft
under grant number DFG-MA 4967/1-1.



\bibliography{HaloModel.FNL.PRD.2.bbl}


\newpage

\begin{widetext}

\appendix


\section{Some results concerning the primordial Skewness}\label{app:skew}

In this appendix we give some relations concerning the skewness of the
density field. The ensemble average of the cube of the density field
using Fourier space representation can be written:
\ba 
\left<\delta^3_M(x)\right>
& = & 
\int \frac{\dk_1}{(2\pi)^3}
\frac{\dk_2}{(2\pi)^3} \frac{\dk_3}{(2\pi)^3}W(k_1,M)W(k_2,M)W(k_3,M)
\left<\delta(\bk_1)\delta(\bk_2)\delta(\bk_3)\right>
{\rm e}^{-i(\bk_1+\bk_2+\bk_3)\cdot\bx}\ .
\ea 
For a Gaussian field the average of the three Fourier modes gives
zero, but for the local model of non-Gaussianity the product is
related to the primordial potential bispectrum. Using Eqns
(\ref{eq:primebi}), (\ref{eq:potbi}),  and (\ref{eq:alpha}), we arrive
at,
\ba 
\left<\delta^3_M(x)\right> 
& = & 
\int \frac{\dk_1}{(2\pi)^3}
\frac{\dk_2}{(2\pi)^3} \frac{\dk_3}{(2\pi)^3}
W(k_1,M)W(k_2,M)W(k_3,M)
\alpha(k_1)\alpha(k_2)\alpha(k_3)\nn\\
& & \times \left<\phiNG(\bk_1)\phiNG(\bk_2)\phiNG(\bk_3)\right>
\exp[-i(\bk_1+\bk_2+\bk_3)\cdot\bx] 
\nn \\
& =& 
\int \frac{\dk_1}{(2\pi)^3}
\frac{\dk_2}{(2\pi)^3} \frac{\dk_3}{(2\pi)^3}
W(k_1,M)W(k_2,M)W(k_3,M)
\alpha(k_1)\alpha(k_2)\alpha(k_3)\
\nn\\
& & \times B_{\phiNG}(\bk_1,\bk_2,\bk_3)(2\pi)^3\delta(\bk_1+\bk_2+\bk_3)
\exp[-i(\bk_1+\bk_2+\bk_3)\cdot\bx] 
\nn \\
& =& 
\int \frac{\dk_1}{(2\pi)^3}
\frac{\dk_2}{(2\pi)^3} \frac{\dk_3}{(2\pi)^3}
W(k_1,M)W(k_2,M)W(k_3,M)
\alpha(k_1)\alpha(k_2)\alpha(k_3)\
\nn\\
& & \times 2\FNL\left[\PphiG(k_1)\PphiG(k_2)+2\cyc\right](2\pi)^3\delta(\bk_1+\bk_2+\bk_3)
\exp[-i(\bk_1+\bk_2+\bk_3)\cdot\bx] \ .
\ea
Computing the delta function integrals separately for each of the
three terms in square brackets and defining \mbox{$k_3^2\equiv
  k_1^2+k_2^2+2k_1k_2\mu$}, we then find that the skewness ($S_3\equiv
\left<\delta^3\right>/\left<\delta^2\right>^2$) has the form:
\be
S_3\sigma^4= 3\FNL \int
\frac{\dk_1}{(2\pi)^3}\frac{\dk_2}{(2\pi)^3}
\alpha(k_1)\PphiG(k_1) 
\alpha(k_2)\PphiG(k_2)
\int_{-1}^{1}d\mu \alpha(k_3) \left[ W(k_1,M)W(k_2,M)W(k_3,M)\right]\ .
\ee

We are also interested in computing the derivatives of the skewness
with respect to the mass variance $\sigma$ up to second order:
\ba 
\frac{d(S_3\sigma^4)}{d\log\sigma} & = & 
\frac{1}{3}\frac{d\log M}{d\log\sigma}\frac{d(S_3\sigma^4)}{d\log R}\ ;\\
\frac{d^2(S_3\sigma)}{d(\log\sigma)^2} 
& = & 
\frac{1}{3}\frac{d^2\log M}{d(\log\sigma)^2}
\frac{d(S_3\sigma^4)}{d\log R}+
\frac{1}{9}
\left(\frac{d\log M}{d\log\sigma}\right)^2
\frac{d^2(S_3\sigma^4)}{d\log R^2}
\ , \ea
where we used the fact that $d\log M/d\log R=3$. The required
auxiliary functions are:
\ba 
\frac{d(S_3\sigma^4)}{d\log R} & = & 3\FNL \int
\frac{\dk_1}{(2\pi)^3}\frac{\dk_2}{(2\pi)^3}
\alpha(k_1)\PphiG(k_1) 
\alpha(k_2)\PphiG(k_2)
\int_{-1}^{1}d\mu \alpha(k_3) \nn \\
& & \times \ 
\left[\frac{}{} 
 W'_1W_2W_3+
 W_1W_2'W_3+
 W_1W_2W_3'
\right] \ ; \\
\frac{d^2(S_3\sigma^4)}{d\log R^2} & = & 3\FNL \int
\frac{\dk_1}{(2\pi)^3}\frac{\dk_2}{(2\pi)^3}
\alpha(k_1)\PphiG(k_1) 
\alpha(k_2)\PphiG(k_2)
\int_{-1}^{1}d\mu \alpha(k_3) \nn \\
& & \times \ 
\left[\frac{}{} 
 W''_1W_2W_3+ W_1W_2''W_3+ W_1W_2W_3'' + 2W'_1W_2'W_3+ 2W_1'W_2W_3'+ 2W_1W_2'W_3'\right] \ ;\\
\frac{d\log\sigma^2(R)}{d\log M} & = & \frac{2}{3\sigma^2(R)}\int
\frac{\dk_1}{(2\pi)^3} P_{\rm Lin}(k_1) W_1W_1' \ ; \label{eq:sigdiff} \\
\frac{d^2\log\sigma^2(R)}{d\log M^2} 
& = & \frac{2}{9\sigma^2(R)} 
\int \frac{\dk_1}{(2\pi)^3} P_{\rm Lin}(k_1) \left[W_1'^2+W_1W_1''\right] 
-\left[\frac{d\log\sigma^2(R)}{d\log M}\right]^2\ ,
\ea
where we have introduced the notation $W_i\equiv W(k_iR)$, $W_i'
\equiv dW(k_iR)/d\log (k_iR)$ and $W_i'' \equiv d^2W(k_iR)/d\log (k_iR)^2$.  For
a real-space top hat filter function we have:
\ba 
W_{\rm TH}(y) 
& = & \frac{3}{y^3}\left[\sin y - y \cos y\right] \ ; \\
W_{\rm TH}'(y)
& = & \frac{3}{y^3}\left[(y^2-3)\sin y + 3y \cos y\right] \ ; \\
W_{\rm TH}''(y)
& = & \frac{3}{y^3}\left[(9-4y^2)\sin y + y (y^2-9) \cos y\right] \ .
\ea
Finally, some relationships that will be of use for calculating the
halo mass function and also the scale-independent contribution to the
halo bias are:
\ba 
\frac{d\log (\sigma S_3)}{d\log\nu} & = &
\frac{1}{\sigma^3}
\left[\sigma^4S_3-\frac{d(\sigma^4S_3)}{d\log\sigma}\right]\ ; \\
\frac{d^2\log (\sigma S_3)}{d\log\nu^2} & = &
\frac{1}{\sigma^3}
\left[3\sigma^4S_3-4\frac{d(\sigma^4S_3)}{d\log\sigma}+
\frac{d^2(\sigma^4S_3)}{d\log\sigma}^2\right] \ .
\ea


\section{A possible solution to the problem of excess power 
on large scales in the halo model}\label{sec:RPTHM}


Several authors have pointed out that the halo model fails to
reproduce the correct clustering statistics on large scales
\cite{Smithetal2003,CooraySheth2002,CrocceScoccimarro2008,Hamausetal2010}. This
arises because the 1-Halo term approaches a constant value
significantly in excess of the shot noise for dark matter particles in
simulations: e.g.
\be 
\lim_{k\rightarrow0} P_{\1H}(k) =
\frac{1}{\rhob^2}\int_{0}^{\infty} dM n(M) M^2 \gg \frac{1}{\nbar_{\delta}} ,
\ee
where $\nbar_{\delta}$ is the number density of dark matter
particles. A more recent criticism of the halo model, in the context
of renormalized perturbation theory (RPT), was given by
\citep{CrocceScoccimarro2008}. We now show how halo exclusion may
resolve this issue.

Consider the RPT formulation of the nonlinear matter power spectrum in
perturbation theory \cite{CrocceScoccimarro2008}:
\be P_{\rm RPT}(k)=G^2(k)P_{\Lin}(k)+P_{\rm MC}(k) \ , \ee
where $G(k)$ is the nonlinear density propagator, which informs us of
how a density mode decorrelates from its initial state; and the term
$P_{\rm MC}$ informs us of the power gained by a single mode from
coupling with all other modes. \citep{CrocceScoccimarro2008} make the
analogy $P_{\rm 2H}\rightarrow G^2(k)P_{\Lin}$ and $P_{\rm
  1H}\rightarrow P_{\rm MC}(k)$. They then measure the quantity
$P-G^2(k)P_{\Lin}$ from simulations and compare the result with the
theoretical predictions for $P_{\rm MC}\propto k^{4}$ and $P_{\rm
  1H}\propto {\rm const}$.  The simulations show a $k^{4}$ slope, and
hence they conclude that the halo model fails on large scales.

Let us now reexamine this issue in the context of halo
exclusion. Firstly, consider again the halo center clustering of
haloes of masses $M_1$ and $M_2$, if we treat this as in
\cite{Smithetal2007} and expand the halo density as a local function
of the nonlinear dark matter density then we have:
\be P^{\rm \hc\hc}_{\cent}=b_1(M_1)b_1(M_2)P_{\rm NL}(k|R)+ {\mathcal
  O}(b_2,b_3,\dots) \ .  \ee
Consider now the very large-scale limit of the halo model, then we
have $U\rightarrow1$ as $k\rightarrow0$. Recalling the conditions
\Eqn{eq:cond1}, \Eqn{eq:cond2} and \Eqn{eq:cond3}, the halo model then
reduces to,
\be 
P_{\rm HM}(k)=P_{\rm NL}(k|R)-P_{\2H}^{\rm exc}(\bk)+P_{\1H}(k) \ .
\ee
If we now insert the RPT expansion for the large-scale smoothed
nonlinear matter power spectrum, then we have:
\be 
P_{\rm HM}(k)=|W(k|R)|^2 \left[G^2(k)P_{\Lin}(k)+P_{\rm MC}(k)\right] 
-P_{\2H}^{\rm exc}(\bk)+P_{\1H}(k) \ .
\ee
On rearranging the above equation, one finds
\be P_{\rm HM}(k)-|W(k|R)|^2 G^2(k)P_{\Lin}(k) = 
|W(k|R)|^2 P_{\rm MC}(k) - P_{\2H}^{\rm exc}(\bk)+P_{\1H}(k)  \ . \ee
In the limit $k\rightarrow0$, $W(k|R)\rightarrow 1 $. The right hand
side of the above equation can be consistent with the simulation
results of \citep{CrocceScoccimarro2008}, if and only if $P_{\1H}-
P_{\2H}^{\rm exc}\rightarrow\epsilon$, where $\epsilon$ is a
sufficiently small quantity that the right side of the above equation
decays as $\propto k^{4}$ for all observable scales of
interest. Using \Eqn{eq:2HExcLS} we may explicitly write $\epsilon$
as:
\be \epsilon \equiv \frac{1}{\rhob^2}\int dM n(M)M^2
-\prod_{i=1}^{2}\left\{\frac{1}{\rhob} \int dM_i n(M_i) M_i\right\}
V(r_1+r_2) \left[1+b(M_1)b(M_2)\bar{\xi}_{\rm NL}(r_1+r_2)\right] 
\label{eq:epsi}\ .\ee
On taking the relation between halo volume and mass to be
\mbox{$M_{\rm vir}=4\pi r_{\rm vir}^3\Delta\rhob/3$}, where $\Delta$
is an overdensity threshold that defines the halo today,
e.g. $\Delta\sim200$, then we may expand the exclusion volume in
powers of the halo mass,
\be
V(r_1+r_2)=\frac{1}{\rhob\Delta}\left[M_1+3M_1^{2/3}M_2^{1/3}
+3M_1^{1/3}M_2^{2/3}+M_2\right]\ .\ee
On inserting this into the second term on the left side of
\Eqn{eq:epsi} we find,
\be 
\rightarrow \frac{1}{\rhob^3\Delta} \int dM_1 n(M_1) M_1 \int dM_2 n(M_2) M_2
\left[2M_1+6M_1^{2/3}M_2^{1/3}\right]
\left[1+b(M_1)b(M_2)\bar{\xi}_{\rm NL}(r_1+r_2)\right]\ .
\ee
Let us consider two possible limits of the above expression: 

\noindent $\bullet$ $\overline{\xi}_{\rm NL}\ll 1$: then we have
\ba
\rightarrow && 
\frac{1}{\rhob^3\Delta} \int dM_1 n(M_1) M_1 \int dM_2 n(M_2) M_2
\left[2M_1+6M_1^{2/3}M_2^{1/3}\right]\nn \\
& = & \frac{2}{\rhob^2\Delta} \int dM_1 n(M_1) M_1^2 +
\frac{6}{\rhob^3\Delta} \int dM_1 n(M_1) M_1^{5/3} \int dM_2 n(M_2) M_2^{4/3}\ .
\ea

\noindent $\bullet$ $\overline{\xi}_{\rm NL}\gg 1$: then we have
\ba
\rightarrow && 
\frac{1}{\rhob^3\Delta} \int dM_1 n(M_1) M_1 b(M_1)  \int dM_2 n(M_2) M_2  b(M_2)
\left[2M_1+6M_1^{2/3}M_2^{1/3}\right]\bar{\xi}_{\rm NL}(r_1+r_2) \ .\label{eq:thisone}
\ea
To get an estimate of this, suppose that the first term in
\Eqn{eq:thisone} dominates over the second and that \mbox{${\rm
    max}\left[\overline{\xi}_{\rm NL}(r_1+r_2|R)\right]\sim f\Delta$},
then we would have 
\be \rightarrow\ \ \epsilon \approx \frac{1}{\rhob^2}\int dM n(M) M^2 
\left[1- 2f b(M)\right] \ .
\ee
We thus see that the correction can be very close to the order of the
resulting large-scale power for the 1-Halo term, and depends strictly
on the quantity $1-2fb(M)$. We note that the case $f=1$ would mean
$\overline{\xi}\sim\Delta$, and that this would be unrealistically
large. This may be argued in the following way: $\Delta$ is the volume
average density at the viral radius, whereas
$b_1(M_1)b_1(M_2)\overline{\xi}_{\rm NL}(r_1+r_2|r)$ is the volume
average correlation function of halo centres smoothed on the scale
$R$, for separations $r_1+r_2$, and excluding the points within the
same halo, thus we conclude that $f<1$. In fact we would need ${\rm
  max}\left[\overline{\xi}_{\rm NL}(r_1+r_2|R)\right]<\Delta/2b(M)$.

As we show in \Fig{fig:GaussPow}, exact evaluation of the 1-Halo and
the 2-Halo exclusion terms on large scales produces two quantities
that almost exactly cancel. We therefore forward halo exclusion as a
risible solution to the problem of excess large-scale power in the
halo model.

As this paper was being submitted to the {\tt arXiv}, a paper
discussing an alternative solution to resolving the break down of the
Halo Model on large scales was proposed by
\citep{ValageasNishimichi2010}.

\end{widetext}


\end{document}